\documentclass[aps,prd,onecolumn,showpacs,groupedaddress,nofootinbib]{revtex4}  
\usepackage{graphicx}  
\usepackage{dcolumn}   
\usepackage{bm}        
\usepackage{amssymb}   
\usepackage{amsmath}
\usepackage{subfigure}
\usepackage{multirow}
\usepackage{color}
\usepackage{xfrac}

\DeclareMathOperator{\Tr}{Tr}
\DeclareMathOperator{\Dirac}{\delta_{\text{D}}}

\begin{document}

% Journal commands
\newcommand\aap{{A\&A}} % Astronomy and Astrophysics
\newcommand\mnras{{MNRAS}} % Monthly notices of the Royal Astronomical Society
\newcommand\apjl{{Astrophys. J., Letters}} % Astrophysical Journal, Letters
\newcommand\apjs{{Astrophys. J., Supplement}} % Astrophysical Journal, Supplement
\newcommand\jcap{{Journal of Cosmology and Astroparticle Physics}} % Journal of Cosmology and Astroparticle Physics
\newcommand\pasj{{Publications of the Astronomical Society of Japan}} % Publications of the Astronomical Society of Japan

\title{3-dimensional spherical analyses of cosmological spectroscopic surveys}

\author{Andrina Nicola} 
\email{andrina.nicola@phys.ethz.ch} 
\affiliation{Institute for Astronomy, Department of Physics, ETH Zurich, Wolfgang-Pauli-Strasse 27, CH-8093 Zurich, Switzerland}
\author{Alexandre Refregier} 
\affiliation{Institute for Astronomy, Department of Physics, ETH Zurich, Wolfgang-Pauli-Strasse 27, CH-8093 Zurich, Switzerland}
\author{Adam Amara} 
\affiliation{Institute for Astronomy, Department of Physics, ETH Zurich, Wolfgang-Pauli-Strasse 27, CH-8093 Zurich, Switzerland}
\author{Aseem Paranjape} 
\affiliation{Institute for Astronomy, Department of Physics, ETH Zurich, Wolfgang-Pauli-Strasse 27, CH-8093 Zurich, Switzerland}

\vskip 0.25cm

\begin{abstract}

Spectroscopic redshift surveys offer great prospects for constraining the dark sector in cosmology. Future surveys will however be both deep and wide and will thus require an analysis in 3-dimensional spherical geometry. We review and compare several methods which have been proposed in the literature for this purpose, focusing in particular on implementations of the spherical harmonic tomography (SHT) power spectrum $C^{i j}_{l}$ and the spherical Fourier Bessel (SFB) power spectrum $C_{l} (k, k')$. Using a Fisher analysis, we compare the forecasted constraints on cosmological parameters using these statistics. These constraints typically rely on approximations such as the Limber approximation and make specific choices in the numerical implementation of each statistic. Using a series of toy models, we explore the applicability of these approximations and study the sensitivity of the SHT and SFB statistics to the details of their implementation. In particular, we show that overlapping redshift bins may improve cosmological constraints using the SHT statistic when the number of bins is small, and that the SFB constraints are quite robust to changes in the assumed distance-redshift relation. We also find that the SHT can be tailored to be more sensitive to modes at redshifts close to the survey boundary, while the SFB appears better suited to capture information beyond the smooth shape of the power spectrum. In this context, we discuss the pros and cons of the different techniques and their impact on the design and analysis of future wide field spectroscopic surveys.

\end{abstract}


\maketitle

\section{\label{sec:Introduction} Introduction}

Constraining the nature and properties of dark energy and dark matter are amongst the most intriguing tasks of current physics. Spectroscopic galaxy redshift surveys offer a way to probe the matter distribution at low redshift which is strongly affected by the properties of the dark sector. Upcoming spectroscopic clustering surveys like DESI \cite{2013arXiv1308.0847L}, HETDEX \cite{2008ASPC..399..115H} and PFS \cite{2014PASJ...66R...1T}, are therefore amongst the most promising tools to achieve these tasks.
As opposed to the CMB which can be analysed through 2-dimensional maps on the sky, galaxy surveys are inherently 3-dimensional, making their analysis more complex.  Depending on the galaxy survey geometry, different analysis methods have thus been proposed.

For surveys with limited angular sky coverage, the sky can be approximated as flat. Therefore the clustering of galaxies can be analysed in 3 dimensional Cartesian coordinates by means of the spatial correlation function $\xi \left( r \right )$ or through its Fourier counterpart, the Cartesian power spectrum $P(k, r)$.  

In recent years, galaxy redshift surveys have become both wider and deeper, leading us to investigate analysis methods other than $P(k, r)$, which do not rely on the flat-sky approximation and which facilitate combination of galaxy clustering data with other cosmological probes.  A statistic which naturally incorporates the curvature of the sky is the spherical harmonic tomography (SHT) power spectrum $C^{i j}_{l}$, the spherical harmonic transform of the angular correlation function at redshifts $z_{i}$ (for theoretical studies see e.g. \cite{2011PhRvD..84f3505B, 2011PhRvD..84d3516C, 2014JCAP...01..042D, 2012MNRAS.427.1891A} and for application to data see e.g. \cite{2013MNRAS.428.3487H, Ho:2012aa}). The 3-dimensional information can partly be retrieved from this tomographic analysis by performing the spherical harmonics decomposition at a number of different redshifts. Tomographic analyses of the matter overdensity field require subdivision of data into bins, since a finite redshift resolution is needed to compute angular correlations in practice.

Another common way to analyse the 3-dimensional matter overdensity field in spherical geometry, which has been applied to galaxy redshift surveys (e.g. \cite{1995MNRAS.275..483H, 2012Rassat}), weak lensing (e.g. \cite{2003MNRAS.343.1327H, 2005PhRvD..72b3516C}) and the integrated Sachs-Wolfe effect \cite{2012MNRAS.422.2341S}, is to measure its spherical Fourier transform. The result is the 3-dimensional spherical Fourier Bessel (SFB) power spectrum $C_{l} (k, k')$ where the angular dependence is encoded in the multipole $l$ and the radial dependence in the wave vector $k$. This statistic allows us to retrieve the clustering information without having to adopt the flat sky approximation or the need for redshift  binning.  

Recently, the spherical harmonic tomography power spectrum has been compared to $P(k, r)$, showing that the two methods yield consistent results \cite{2012MNRAS.427.1891A} and both these methods have been employed to investigate the complementarity of weak lensing and galaxy redshift surveys (see e.g. \cite{2013arXiv1307.8062K, 2013arXiv1308.6070D}).

With the aforementioned galaxy redshift surveys under development, it becomes increasingly important to further test and compare the applicability of these statistics to survey requirements. In this paper, we compare the two spherical-sky statistics, i.e. the SHT power spectrum and the SFB power spectrum using a Fisher analysis. We study the sensitivity of these statistics to the detail of their implementation, placing particular emphasis on the advantages and disadvantages of each method, some of which we illustrate with simplified toy models.

This paper is organised as follows. In Section \ref{sec:Baseline} we summarise our comparison baseline model. In Sections \ref{sec:3D_PowerSpectra} and \ref{sec:FisherMatrices} we review 3-dimensional spherical analyses of the matter overdensity field as well as Fisher matrix forecasting techniques and present applications to the SFB power spectrum. In Section \ref{sec:Results} we present a comparison of the spherical harmonic tomography and the SFB power spectrum. We conclude in Section \ref{sec:Conclusions}. Derivations and discussion of employed toy models are deferred to the Appendix.

\section{\label{sec:Baseline} Comparison baseline}

In this work, we consider a $\text{wCDM}$ cosmological model in the framework of general relativity specified by the set of 7 cosmological parameters $\boldsymbol{\theta} = (h, \Omega_{\text{m}}, \Omega_{\Lambda}, w_{0}, w_{\text{a}}, n_{\text{s}}, \sigma_{8})$, where we fix the baryon density $\Omega_{\text{b}} = 0.045$. This model allows for a dynamical evolution of dark energy as well as curvature and is characterised by 7 parameters: the mean fractional matter density $\Omega_{\text{m}}$, the fractional density of dark energy $\Omega_\Lambda$, the Hubble constant $H_0=100h\,$km/s/Mpc, the r.m.s. of matter fluctuations $\sigma_{8}$ in spheres of comoving radius $8h^{-1}$Mpc, the scalar spectral index $n_{\text{s}}$ and two parameters $w_0$ and $w_{\text{a}}$ that characterise the equation of state of dark energy (\cite{2003PhRvL..90i1301L, 2001IJMPD..10..213C})
\begin{equation}
w(a) = w_{0} + (1- a) w_{\text{a}}
\end{equation}
We choose fiducial values $\boldsymbol{\theta}_{\text{fid}} = (0.7, 0.3, 0.69, -0.95, 0, 1.0, 0.8)$, which are consistent with the recent results by WMAP 9 \cite{2013ApJS..208...19H}. In all calculations we fix $w_{\text{a}} = 0$.

\begin{figure}
\includegraphics[scale=0.5]{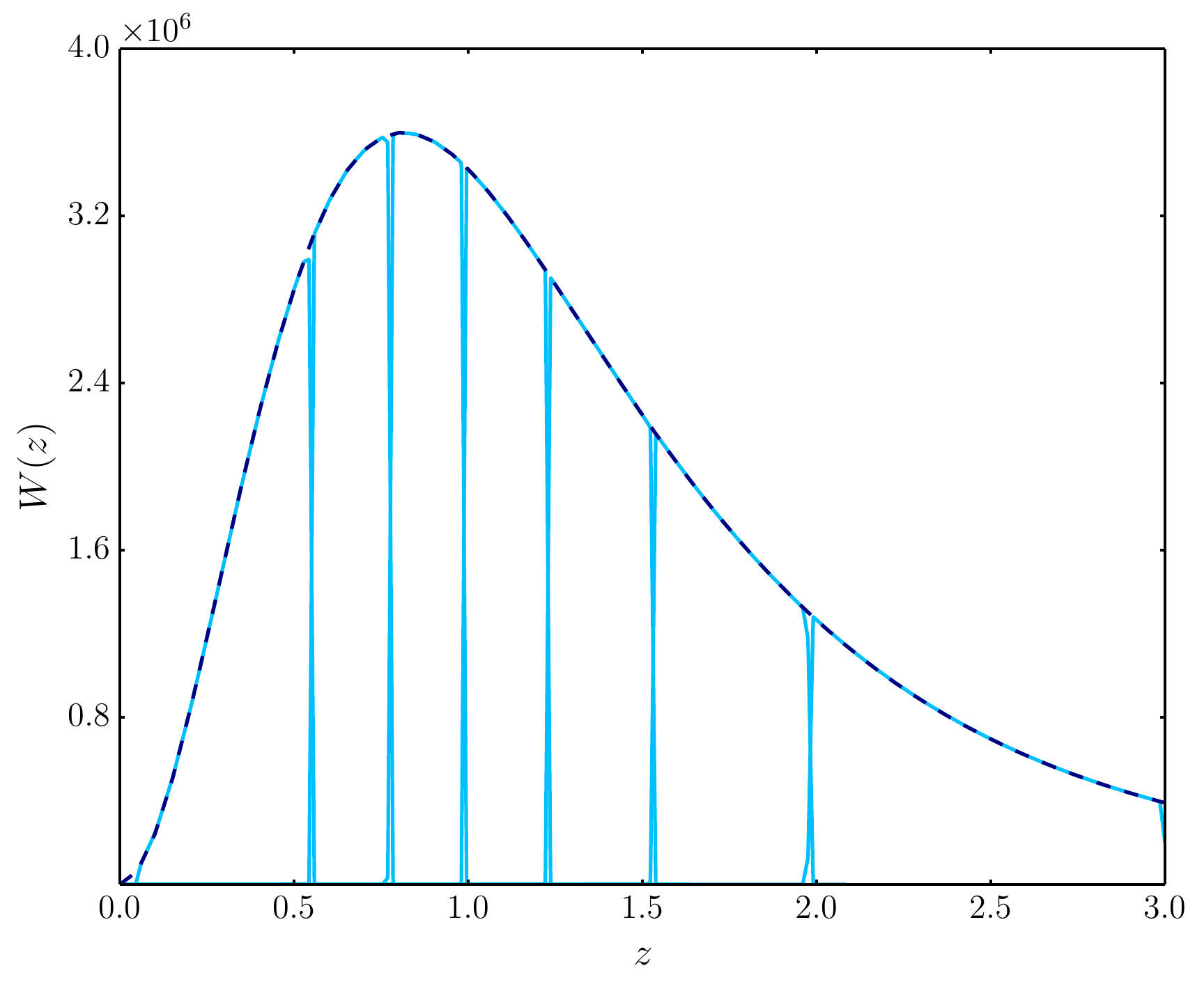}
\caption{\label{fig:selection_func} Survey window function with mean redshift $\bar{z} = 1.2$ and constant galaxy volume density $n_{0} = 10^{-3} \left ( h^{-1} \text{Mpc} \right)^{-3}$. This corresponds to a galaxy surface density of $n_{\text{A}} = 0.49 \; \text{arcmin}^{-2}$.}
\end{figure}

We define the baseline survey in terms of total surveyed volume $V$ and number of detected galaxies $N_{\text{gal}}$ by expressing the galaxy number density as \cite{2013PhRvD..88b3502Y}
\begin{equation}
n \left( r \right ) = n_{0} \; \phi \left( r \right ) = \frac{N_{\text{gal}}}{V} \; \phi \left( r \right )
\label{eq:SFB_NumberDensity}
\end{equation}
where $r$ is the comoving distance and $\phi \left( r \right )$ defines the radial survey selection function with normalisation $\int \text{d}^{3} \textbf{r} \; \phi(r) = V$ as in \cite{2012Rassat, 2013PhRvD..88b3502Y}. The normalisation condition defines the window function in redshift $W(z)$ \cite{2013PhRvD..88b3502Y}. For incomplete sky-coverage, parametrised by the fraction of sky covered in the survey $f_{\text{sky}}$, it becomes
\begin{equation}
\int \text{d} r \; r^{2} \phi(r) = \int \text{d} z \; \frac{c}{H(z)} r^{2} \phi(r) = \int \text{d} z \; W(z) = \frac{V}{4 \pi f_{\text{sky}}}
\end{equation}

For all surveys considered, we assume fractional sky coverage $f_{\text{sky}} = 0.125$, which represents a lower limit to the sky fraction covered by future surveys, and a radial selection function given by
\begin{equation}
\phi \left( r \right ) = e^{- \left (\frac{r}{r_{0}} \right )^{2}}
\label{eq:SFB_SelectionFunc}
\end{equation}
which yields a survey volume of $V = \pi^{\frac{3}{2}} r_{0}^{3}$. This choice is motivated by the fact that it allows for the analytical computation of SFB power spectra in the absence of evolution, which is useful to test the full results. In order to match window functions characteristic of upcoming galaxy redshift surveys, we set $r_{0} = r \left (z = 1 \right ) = 2354 \; h^{-1} \text{Mpc}$, as computed in our fiducial cosmological model.

Fig.~\ref{fig:selection_func} shows the angular galaxy density as a function of redshift for a volume density of $n_{0} = 10^{-3} \left ( h^{-1} \text{Mpc} \right)^{-3}$; it further illustrates the binning scheme adopted in the tomographic analysis. Each bin configuration is chosen by requiring the same number of galaxies in each redshift bin. Our baseline choices are summarised in Table \ref{tab:baseline}. For comparison and future reference we show recent choices made in the literature in Table \ref{tab:papers_choices}. This table highlights the breadth of possible choices to make when analysing a galaxy redshift survey. The most important ones include:

\begin{enumerate}

\item Statistic used: spherical harmonic tomography power spectrum, spherical Fourier Bessel power spectrum or Cartesian power spectrum

\item Physical effects included: redshift space distortions, relativistic corrections (see e.g. \cite{2011PhRvD..84f3505B, 2013PhRvD..88b3502Y})

\item Implementation scheme: examples include simplifying assumptions like the Limber approximation, Fisher matrix computation method or number of cosmological parameters considered

\end{enumerate}

From Table \ref{tab:papers_choices} we see that the choices made differ considerably. One of the aims of this paper is therefore to investigate how much parameter forecasts and constraints are influenced by some of these choices.

\begin{table}
\caption{Baseline specification} \label{tab:baseline}
\begin{center}
\begin{tabular}{>{\centering}m{2.5cm}|>{\centering}m{7cm}|>{\centering}m{8cm}@{}m{0pt}@{}} \hline \hline
                                 
Survey & 
\multicolumn{2}{>{\centering}m{15cm}}{Sky coverage: $f_{\text{sky}} = 0.125$ \\
Selection function: $\phi \left( r \right )$ as in Eq.~\ref{eq:SFB_SelectionFunc} with $r_{0} = 2354 \; h^{-1} \text{Mpc}$\\
Galaxy volume density: $n_{0} = 10^{-3} \left ( h^{-1} \text{Mpc} \right)^{-3}$ \\
Galaxy surface density: $n_{\text{A}} =  0.49 \; \text{arcmin}^{-2}$ \\
Angular scales covered: $l \in [2, 50]$} &

\tabularnewline \hline
Model & 
\multicolumn{2}{>{\centering}m{15cm}}{Cosmological parameters: $\boldsymbol{\theta} = (h, \Omega_{\text{m}}, \Omega_{\Lambda}, w_{0}, n_{\text{s}}, \sigma_{8}) \footnote{We further fix $\Omega_{\text{b}} = 0.045$ and $w_{\text{a}} = 0$.}$ \\
Fiducial values: $\boldsymbol{\theta}_{\text{fid}} = (0.7, 0.3, 0.69, -0.95, 1.0, 0.8)$ \\
Prior: none \\
Galaxy bias: $b \left (k, r \right ) = 1$ \\
Redshift space distortions: yes \\
Relativistic corrections: no} &

\tabularnewline \hline

Implementation &
SFB \\[1ex] 
k range: $k \in (0.0007, 0.2) \: h \text{Mpc}^{-1}$ \\
Fisher matrix: diagonal
& 
SHT \\[1ex]
Number of bins: 7 \\
Redshift range: $0.05 \leq z \leq 3.0$ &

\tabularnewline \hline \hline

\end{tabular}
\end{center}
\end{table} 

\begin{table*}
\caption{\label{tab:papers_choices} Compilation of different implementations used in the literature.}
\begin{ruledtabular}
\begin{tabular}{cccccccc>{\centering\arraybackslash}p{2.5cm}}
& Paper & Statistic  & RSD & \parbox{1.7cm}{Rel. \\ corrections} & Bias &  \parbox{1.3cm}{Limber \\ approx.} & $N_{\text{param}}$ & Prior \\[1.5ex] \hline
 \multirow{8}{1.5cm}{Fisher analysis} &Gaztanaga et al., 2012 \cite{2012MNRAS.422.2904G}&$C_{l}^{ij}, P(k)$ &yes\footnote{Transverse modes with $C_{l}^{ij}$. RSDs taken into account for the radial modes in each redshift bin using $P(k)$.}&no&yes&yes&8&Planck + SN-II \\
 &de Putter et al., 2013 \cite{2013arXiv1308.6070D}&$C_{l}^{ij}, P(k)$
 &$\text{yes}^{a}$&no&yes&yes&9&Planck\\
 &Cai et al., 2012 \cite{2012MNRAS.422.1045C}&$C_{l}^{ij}, P(k)$
 &$\text{yes}^{a}$&no&yes&yes&6&CMB p. s.\footnote{Assume primordial CMB power spectrum known.}\\
 &Kirk et al., 2013 \cite{2013arXiv1307.8062K}&$C_{l}^{ij}$&yes&no&yes&no&7&flat\\
 &Font-Ribera et al., 2013 \cite{Font-Ribera:2014aa}&$C_{l}^{ij}, P(k)$&$\text{yes}^{a}$&no&yes&yes&8&Planck\\ 
 &Di Dio et al.,2014 \cite{2014JCAP...01..042D}&$C_{l}^{ij}$&yes&yes&no&no&5&none \\ 
 & This work &$C_{l}^{ij}, C_{l} (k, k')$&yes&no&no&no&6&none \\ \hline
\multirow{5}{1.5cm}{General analysis}  &Bonvin et al., 2011 \cite{2011PhRvD..84f3505B}&$C_{l}^{ij}$&yes&yes&no&no& - & - \\
 &Challinor et al., 2011 \cite{2011PhRvD..84d3516C}&$C_{l}^{ij}$&yes&yes&yes&no& - & - \\
 &Rassat et al., 2012 \cite{2012Rassat} &$C_{l} (k, k')$& no &no&no&no& - & -\\
 &Yoo et al., 2013 \cite{2013PhRvD..88b3502Y}&$C_{l} (k, k')$
 &yes&yes&no&no& - & -\\
 &Pratten et al., 2013 \cite{2013MNRAS.436.3792P}&$C_{l} (k, k')$
 &yes&no&no&no& - & -\\
\end{tabular}
\end{ruledtabular}
\end{table*}

\section{\label{sec:3D_PowerSpectra} 3-dimensional spherical power spectra}

\subsection{\label{subsec:cart_ps} The Cartesian power spectrum}

A common approach to analyse the observed matter overdensity field $\delta \left ( \textbf{x} , r \right )$ is to expand it into its Cartesian Fourier components $\delta \left (\textbf{k}, r \right ) $, where we use the comoving distance $r$ as a measure of time $t$. The real-space overdensity field is related to its Fourier counterpart through
\begin{equation}
\delta \left ( \textbf{x}, r \right ) = \frac{1}{\left ( 2 \pi \right )^{3}} \int \text{d}^{3} \textbf{k} \; \delta \left (\textbf{k}, r \right ) e^{i \textbf{k} \cdot \textbf{x}}
\label{eq:Fourier_conv}
\end{equation}
The Fourier space correlation function is the Cartesian matter power spectrum $P \left( k, r \right )$ defined by
\begin{equation}
\langle  \delta ( \textbf{k}, r )  \delta ( \textbf{k}', r ) \rangle = \left ( 2 \pi \right )^{3} \Dirac ( \textbf{k} - \textbf{k}'  ) P \left( k, r \right )
\label{eq:powerspec}
\end{equation}
$\langle \rangle$ denotes an ensemble average. We only focus on the linear matter power spectrum which factorises into a time and scale dependent part as
\begin{equation}
P \left( k, r \right ) = D^{2}(r) P_{0}(k)
\end{equation}
where $D(r)$ is the linear growth factor and $P_{0}(k)$ is the power spectrum at redshift $z = 0$. We assume a transfer function as summarised in \cite{1995ApJS..100..281S} and neglect both the effects of baryon oscillations (BAOs) and neutrinos in our baseline configurations. In Section \ref{sec:Results} we investigate the impact of BAOs on our results.

The measurement of this statistic from galaxy redshift surveys bears one complication: the observables in these surveys are the galaxy redshifts $z$ and their angular positions on the sky $\left ( \theta, \phi \right )$. Therefore, in order to compute any 3-dimensional power spectrum from data, the redshift needs to be related to a wave number $\textbf{k}$ through the assumption of a radial distance. This transformation depends on the choice of a fiducial cosmological model. Any 3-dimensional analysis of the matter overdensity field therefore requires the assumption of a cosmological model \cite{2012MNRAS.426.2719R}, prior to testing it.

Another consequence of the fact that radial galaxy distances are only accessible through their redshift, is that the distance estimates will be affected by peculiar galaxy velocities. The comoving galaxy distances $\textbf{s}$ inferred from their redshifts are given by \cite{1987MNRAS.227....1K}
\begin{equation}
\textbf{s} = \textbf{r} + \frac{\textbf{v} \cdot \textbf{n}}{a \; H(a)} 
\label{eq:RSDdist}
\end{equation}
where $\textbf{v}$ is the galaxy velocity due to the linear collapse of overdensities, $H(a)$ is the Hubble parameter and $\textbf{n}$ denotes the line of sight direction. These redshift space distortions (RSDs) lead to an enhancement of the Cartesian matter power spectrum given by \cite{1987MNRAS.227....1K}
\begin{equation}
P_{\text{s}} \left( \mathbf{k}, r \right ) = P \left( k, r \right ) \left (1 + \beta \mu_{k}^{2} \right )^{2}
\label{eq:powerspec_RSD}
\end{equation}
where $\beta = \sfrac{f}{b}$. The quantity $f =  \sfrac{d \ln D(a)}{d \ln a}$ denotes the linear growth rate, $b$ is the galaxy bias, discussed below, and $\mu_{k}$ is the cosine of the angle between the line of sight and the wave vector $\mathbf{k}$. Measuring the power spectrum in redshift space therefore allows us to also estimate the growth rate $f$ of matter perturbations. 

Irrespective of the analysis method, galaxy redshift surveys pose an additional complication. Since galaxies are only expected to form inside the peaks of the overdensity field \cite{1986ApJ...304...15B} and galaxy formation is not completely understood yet, the galaxy overdensity field $\delta_{\text{g}} (k, r)$  is expected to constitute a biased tracer of the underlying dark matter distribution $\delta_{\text{dm}} (k, r)$, i.e. $\delta_{\text{g}} (k, r) = b (k, r) \: \delta_{\text{dm}} (k, r)$. In this paper we assume that galaxies perfectly trace dark matter, which amounts to setting the bias parameter $b(k, r) = 1$. Since in this work we focus on clustering on large scales, where the scale-dependence of galaxy bias is negligible (e.g. \cite{2011MNRAS.415..383M}), we believe that this simplified assumption is appropriate because the statistics we compare will all equally suffer from the problem of bias. An investigation of the effects of scale-dependent bias on our results would be interesting for future work.

\subsection{\label{subsec:Clz} The spherical harmonic tomography power spectrum}

The need for assuming a cosmological model, which arises in 3-dimensional analyses of galaxy clustering, can be circumvented with a tomographic analysis. This amounts to discretising the redshift and analysing the angular dependence of galaxy clustering through the spherical harmonic tomography power spectrum $C^{ij}_{l}$ at a number of different redshifts in order to partly recover the 3-dimensional information. In practice, a galaxy catalogue is analysed by subdividing the galaxies into redshift bins and computing both the auto- and cross-power spectra for all the bins.

\begin{figure}
\includegraphics[scale=0.53]{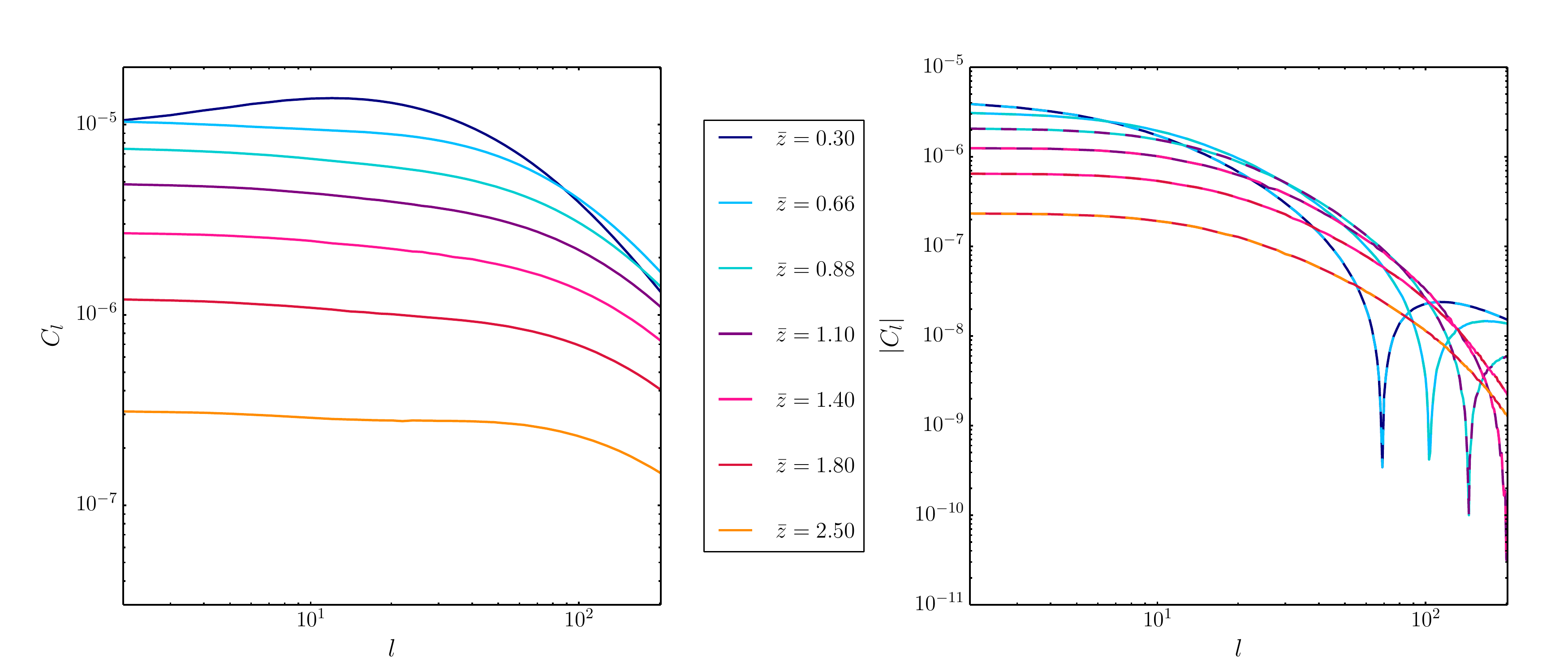}
\caption{\label{fig:distance_measures} Spherical harmonic tomography power spectra $C_{l}^{ij}$ for the 7 redshift bins and survey specified in Sec.~\ref{sec:Baseline}, where $\bar{z}$ denotes the mean redshift of each bin. The left panel shows the auto-power spectra, while the absolute value of the cross-power spectra are shown on the right hand side.}
\label{fig:Cls_SHT}
\end{figure}

Assuming the overdensity field to be statistically isotropic and homogeneous, the spherical harmonic tomography power spectrum including RSDs between redshift bins $i$ and $j$, with radial selection functions $\phi_{i}(r)$ and $\phi_{j}(r)$ respectively, is given by \cite{2007MNRAS.378..852P}
\begin{equation}
C^{ij}_{l} = \frac{2}{\pi} \int \text{d} k k^{2} P_{0} \left(k \right) \left ( W^{i}_{l} \left(k \right) + \beta W_{l}^{i, r} \left(k \right) \right )  \left ( W^{j}_{l} \left(k \right) + \beta W_{l}^{j, r} \left(k \right) \right )
\label{eq:ClRSD}
\end{equation}
where the auto power spectra are obtained for $i = j$ and the cross power spectra for $i \neq j$. The selection functions are normalised i.e. $\int \text{d} r \: \phi_{i}(r) = 1$. Their unnormalised counterparts, the redshift distributions for each bin, are shown in Fig.~\ref{fig:selection_func}. $W^{i}_{l}$ is the real-space window function whereas $W_{l}^{i, r}$ accounts for the corrections due to RSDs; both window functions are defined in terms of the spherical Bessel functions $j_{l}$ as \cite{2007MNRAS.378..852P}
\begin{equation}
W_{l} (k) = \int \text{d} r D(r) \phi_{i}(r) j_{l} \left(k r \right )
\label{eq:Wl}
\end{equation}
\begin{align}
W^{r}_{l} \left(k \right) &= \int \text{d} r D(r) \phi_{i} \left(r \right) \left[ \frac{\left(2 l^{2} + 2 l - 1 \right)}{\left(2 l + 3 \right) \left(2 l - 1 \right)} j_{l} \left(k r \right)  \right. \nonumber \\
&  - \left. \frac{l \left(l-1 \right)}{\left(2 l -1 \right) \left(2 l + 1 \right)} j_{l-2} \left(k r \right) - \frac{\left(l+1 \right) \left(l+2 \right)}{\left(2l+1 \right) \left(2l+3 \right)} j_{l+2} \left(k r \right) \right]
\label{eq:Wl_RSD}
\end{align}
Fig.~\ref{fig:Cls_SHT} shows both the auto and the cross (neighbouring redshift bins) SHT power spectra $C_{l}^{ij}$ as a function of angular scale $l$ for the baseline configuration defined in Sec.~\ref{sec:Baseline}.

The computation of the spherical harmonic tomography power spectrum through Eq.~\ref{eq:ClRSD} can be computationally expensive and it is therefore common to resort to the small angle and wide selection function approximation. At large $l$, the spherical harmonic tomography power spectrum can be approximated through Limber's approximation as \cite{1953ApJ...117..134L}
\begin{equation} 
C^{ij}_{l} \simeq \int \text{d} r \frac{\phi_{i}(r) \phi_{j}(r) }{r ^{2}} D^{2}(r) P_{0} \left (k = \frac{l+\frac{1}{2}}{r} \right ) 
\label{eq:ClLimber}
\end{equation}

\subsection{\label{subsec:SFB} The spherical Fourier Bessel power spectrum}

An alternative way for analysing the galaxy overdensity field is through the spherical Fourier Bessel transform (e.g. \cite{1995MNRAS.275..483H, 2012Rassat, 2013MNRAS.436.3792P, 2013PhRvD..88b3502Y}). The galaxy overdensity field $\delta (\textbf{r})$ can be expanded into its translationally invariant parts i.e. the eigenfunctions of the Laplacian in spherical coordinates. In flat-space these are given by products of spherical Bessel functions and spherical harmonics $Y_{lm}(\theta, \phi )$ which leads to the expansion
\begin{equation}
\delta (\textbf{r}) = \sqrt{\frac{2}{\pi}} \int \text{d} k \sum_{l, m} \delta_{lm}(k) k j_{l} \left (k r \right ) Y_{lm}(\theta, \phi ) 
\label{eq:SFBcoeffs}
\end{equation}
The coefficients $\delta_{lm} (k)$ are given by
\begin{equation}
\delta_{lm} (k) =  \frac{1}{2 \pi^{2}} \sqrt{\frac{2}{\pi}} k i^{l} \int \text{d} r \; r^{2} \int \text{d}^{3} k' \; \delta \left( \mathbf{k'},  r\right ) j_{l} \left (k r \right ) j_{l} \left (k' r \right ) Y^{*}_{lm}(\theta_{k'}, \phi_{k'} ) 
\label{eq:SFBcoeffs_noSIH}
\end{equation}
The SFB power spectrum is defined as the variance of these coefficients as given below, where the last equality holds if the overdensity field $\delta (\textbf{r})$ is statistically isotropic and homogeneous (SIH) \cite{2012Rassat}
\begin{equation}
\langle \delta_{lm} (k) \delta^{*}_{l'm'} (k') \rangle = C_{l} (k, k') \delta_{ll'} \delta_{mm'} = C_{l} (k) \Dirac(k-k') \delta_{ll'} \delta_{mm'}
\end{equation}
Under the SIH assumption we further obtain \cite{2005PhRvD..72b3516C}
\begin{equation}
C_{l}(k) = P \left( k \right )
\label{eq:SFB_PS}
\end{equation}
A similar identity can be obtained in the presence of RSDs in the flat-sky limit and for high radial wavevectors $k$. In this case we approximately obtain
\begin{equation}
C_{l} (k) \simeq P \left( k \right ) \left ( 1 + \beta \right )^{2}
\end{equation}
A derivation of this radialisation in the presence of RSDs can be found in Appendix \ref{ap:Radialisation_SFB}. In cosmology, the SIH condition will usually be violated for two reasons \cite{2012Rassat} (i) the observed fields are generally confined to finite regions of space defined by the survey selection function and (ii) the field $\delta (\textbf{r})$ and the power spectrum $P \left( k, r \right )$ evolve.

In order to compute the observed SFB power spectrum, constraints on radial survey geometry can be imposed through a radial selection function $\phi(r)$, as defined in Sec. \ref{sec:Baseline} \cite{2012Rassat}
\begin{equation}
\delta_{\text{obs}} (\textbf{r}) = \phi(r) \delta (\textbf{r})
\label{eq:f_obs}
\end{equation} 
Accounting for time-evolution as well as the effects of RSDs, the observed SFB power spectrum of the overdensity field $\delta$ becomes \cite{2012Rassat}
\begin{equation}
C_{l} (k, k') = \left (\frac{2}{\pi} \right )^{2} \int \text{d} k'' k''^{2} P_{0}(k'') \left ( W_{l}(k, k'') + W^{r}_{l}(k, k'') \right ) \left ( W_{l}(k', k'') + W^{r}_{l}(k', k'') \right )
\label{eq:SFBobs_RSD}
\end{equation}
$W_{l}(k, k'')$ is the real-space window function whereas $W^{r}_{l}(k, k'')$ accounts for the corrections due to RSDs; they are defined as (\cite{2012Rassat, 2013MNRAS.436.3792P})
\begin{equation}
W_{l} (k, k'') =  \int \text{d} r r^{2} D(r) \phi(r) k j_{l} \left (k r \right ) j_{l} \left (k'' r \right )
\label{eq:Wl_SFB}
\end{equation}
\begin{align}
W^{r}_{l} (k, k'') &=  \int \text{d} r r^{2} \beta \frac{k^{2}}{k''} D(r) \phi(r) \left [ \frac{l^{2}}{(2l+1)^{2}} j_{l-1} \left (k r \right ) j_{l-1} \left (k'' r \right ) \right. \nonumber \\
& - \left. \frac{l(l+1)}{(2l+1)^{2}} \left \{ j_{l-1} \left (k r \right ) j_{l+1} \left (k''r \right ) + j_{l+1} \left (k r \right ) j_{l-1} \left (k''r \right ) \right \} + \frac{(l+1)^{2}}{(2l+1)^{2}} j_{l+1} \left (k r \right ) j_{l+1} \left (k''r \right ) \right ] \label{eq:SFB_RSD} \\ 
& + \int \text{d} r r^{2} \beta \frac{k}{k''} D(r) \frac{\text{d} \phi(r)}{\text{d} r} \left ( \frac{l}{(2l+1)}  j_{l} \left (k r \right ) j_{l-1} \left (k''r \right ) - \frac{(l+1)}{(2l+1)}  j_{l} \left (k r \right ) j_{l+1} \left (k'' r \right ) \right ) \nonumber 
\end{align}
Eq.~\ref{eq:SFB_RSD} allows for a time-dependence of the overdensity field $\delta (\textbf{k}, r)$, since the survey selection functions tend to be broad in redshift as opposed to the redshift bins in Section \ref{subsec:Clz}. 

\begin{figure}
\begin{center}
\subfigure{\includegraphics[width=0.49\textwidth]{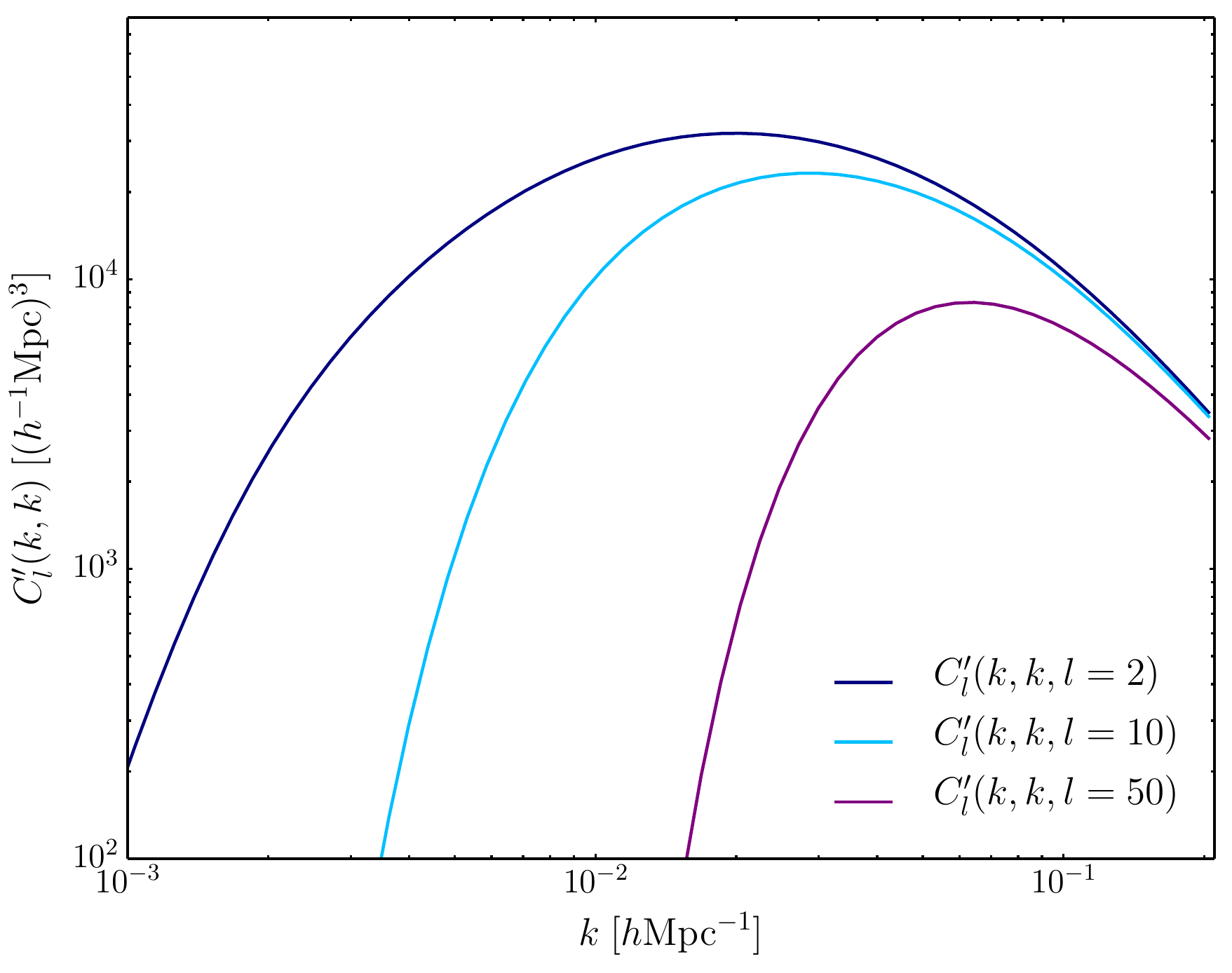}}
\subfigure{\includegraphics[width=0.49\textwidth]{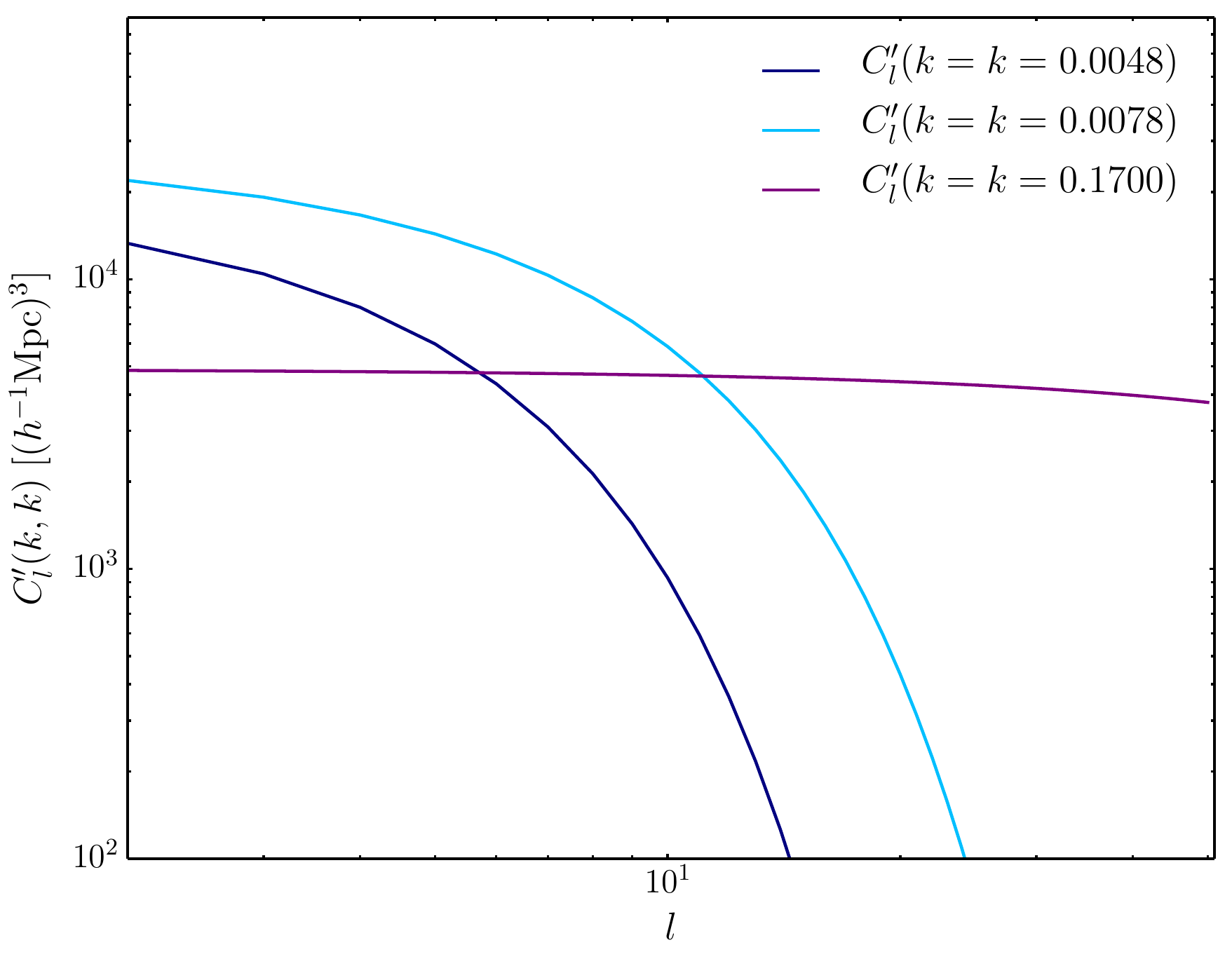}} 
\caption{The SFB auto power spectrum $C'_{l} (k, k)$ as a function of wave vector $k$ and angular scale $l$.} 
\label{fig:Cls_SFB}
\end{center}
\end{figure}

Fig.~\ref{fig:Cls_SFB} shows the SFB power spectrum $C'_{l} (k, k) = C_{l} (k, k) \left( \sfrac{r_{0}}{2 \sqrt{2 \pi}} \right)^{-1}$, both as a function of angular scale $l$ and wave vector $k$ for the selection function defined in Eq.~\ref{eq:SFB_SelectionFunc}. The normalisation follows \cite{2013PhRvD..88b3502Y} and facilitates comparison of the SFB with the Cartesian power spectrum $P \left( k, r \right )$.

Just as for the spherical harmonic tomography power spectrum it is useful to obtain approximations to Eq.~\ref{eq:SFBobs_RSD}. In \cite{2013PhRvD..88b3502Y} it is shown that in the limit of large angular multipoles $l$, the SFB power spectrum, neglecting RSDs, can be approximated by \cite{2013PhRvD..88b3502Y}
\begin{equation}
C_{l} (k, k') \simeq \phi^{2} \left ( \frac{l+\frac{1}{2}}{k} \right ) D^{2} \left (\frac{l+\frac{1}{2}}{k} \right ) P_{0}(k)  \Dirac \left (k - k' \right )
\label{eq:SFB_Limber}
\end{equation}
In the cases considered, there always remains a significant difference between Eq.~\ref{eq:SFBobs_RSD} and Eq.~\ref{eq:SFB_Limber}, which is the SFB analogue of Limber's approximation. Nonetheless it proves useful to test the accuracy of the full equation.

\subsubsection{\label{subsubsec:Alt_Distances} The SFB power spectrum for a generic distance-redshift relation}

The spherical Fourier Bessel coefficients are functions of the wave vector $k$, which in turn depends on the measure of separation in real space. It is customary to choose $k$ conjugate to the comoving separation $r$, making the assumption of a cosmological model inevitable when computing the SFB transform. As an alternative, in this paper we additionally compute the SFB power spectrum for two distance-redshift relations which can directly be computed from observable quantities. 

In order to derive an expression for the SFB power spectrum, we assume a generic distance-redshift relation defined as $\tilde r (z)$ where $\tilde r$ is an arbitrary monotonic function of the redshift $z$. 

With this choice of distance-redshift relation the SFB power spectrum reduces to
\begin{equation}
C_{l} (\nu, \nu') = \left (\frac{2}{\pi} \right )^{2} \int \text{d} k'' k''^{2} P_{0}(k'') \left ( W_{l}(\nu, k'') + W^{r}_{l}(\nu, k'') \right ) \left ( W_{l}(\nu', k'') + W^{r}_{l}(\nu', k'') \right )
\label{eq:SFB_dist}
\end{equation}
where $\nu$ denotes the wave vector conjugate to the new separation measure and $\phi'$ is the selection function in this coordinate system. $W_{l}(\nu, k'')$ is the real-space window function whereas $W^{r}_{l}(\nu, k'')$ accounts for the corrections due to RSDs; they are given by
\begin{equation}
W_{l} (\nu, k'') =  \int \text{d} r r^{2} D(r) \phi'(\tilde r) \nu j_{l} \left (\nu \tilde r \right ) j_{l} \left (k'' r \right )
\label{eq:SFB_window_noRSD}
\end{equation}
\begin{align}
W^{r}_{l} (\nu, k'') &=  \int \text{d} r r^{2} \beta \frac{\nu^{2}}{k''} D(r) \phi'(\tilde r) \frac{\text{d} \tilde r}{\text{d} z} \frac{H(z)}{c} \left [ \frac{l^{2}}{(2l+1)^{2}} j_{l-1} \left (\nu \tilde r \right ) j_{l-1} \left (k'' r \right ) \right. \nonumber \\ 
& - \left. \frac{l(l+1)}{(2l+1)^{2}} \left \{ j_{l-1} \left (\nu \tilde r \right ) j_{l+1} \left (k''r \right ) + j_{l+1} \left (\nu \tilde r \right ) j_{l-1} \left (k''r \right ) \right \} + \frac{(l+1)^{2}}{(2l+1)^{2}} j_{l+1} \left (\nu \tilde r \right ) j_{l+1} \left (k''r \right ) \right ] \label{eq:SFB_window_RSD} \\ 
& + \int \text{d} r r^{2} \beta \frac{\nu}{k''} D(r) \frac{\text{d} \phi'}{\text{d} \tilde r} \frac{\text{d} \tilde r}{\text{d} z} \frac{H(z)}{c} \left ( \frac{l}{(2l+1)}  j_{l} \left (\nu \tilde r \right ) j_{l-1} \left (k''r \right ) - \frac{(l+1)}{(2l+1)}  j_{l} \left (\nu \tilde r \right ) j_{l+1} \left (k'' r \right ) \right ) \nonumber 
\end{align}
Since the selection function transforms as a scalar quantity, $\phi' \left ( \tilde r \right )$ is related to the selection function in terms of the comoving distance through $\phi' \left ( \tilde r(z) \right ) = \phi \left (r(z) \right )$. For a derivation of these identities, the reader is referred to Appendix \ref{ap:Alt_Distances}.

As an illustration of the impact of the choice of distance-redshift relation, we consider two alternatives to the comoving distance 
\begin{eqnarray}
&& \tilde r(z) = \frac{c}{H_{0}} z \label{eq:SFB_Distance_meas} \\
&& \tilde r(z) = \frac{c}{H_{0}} \ln(1+z) \nonumber 
\end{eqnarray}
The first is the linear approximation to the comoving distance valid for low redshifts;, while the second is a logarithmic approximation to $r$ chosen to reproduce both its behaviour at low and intermediate redshift. The resulting SFB power spectra are shown in Fig.~\ref{fig:Cls_SFB_dist}. The normalisation again follows \cite{2013PhRvD..88b3502Y}. 

\section{\label{sec:FisherMatrices} Fisher matrices for 3D spherical power spectra}

The Fisher matrix (FM) allows us to forecast the constraints on cosmological parameters obtainable with future surveys under the approximation of Gaussianity (for an overview of Fisher forecasting see e.g. \cite{1997ApJ...480...22T, 2009arXiv0906.0664H} on which this summary is based). This method can be applied to survey optimisation or, as done in this paper, it can be used to assess the constraining power of different data analysis methods. The FM allows for the propagation of uncertainties in the measurement to uncertainties on the model parameters, which here are the parameters of the $\text{wCDM}$ cosmological model. Bayes' theorem allows us to relate the posterior probability distribution $p \left (\boldsymbol \theta \mid \textbf{x} \right )$ around the maximum likelihood (ML) estimator to the data likelihood $L \left (\textbf{x}; \boldsymbol \theta \right )$. The inverse covariance matrix of the posterior distribution is called the Fisher matrix and given by 
\begin{equation}
F_{\alpha \beta} = \langle - \frac{\partial^{2} \ln{L}}{\partial \theta_{\alpha} \partial \theta_{\beta}} \rangle
\label{eq:FisherMat}
\end{equation}
When several parameters are simultaneously estimated from the data, the marginalised uncertainty on each parameter $\theta_{\alpha}$ is bounded by $\Delta \theta_{\alpha} \ge \sqrt{F^{-1}_{\alpha \alpha}}$ \cite{Kendall1973}. The fixed uncertainty, obtained when keeping all parameters except one fixed, is smaller or equal to the former and given by $\Delta \theta_{\alpha} \ge \sfrac{1}{\sqrt{F_{\alpha \alpha}}}$ \cite{Kendall1973}.

\subsection{\label{subsec:FM_Clz} The Fisher matrix for the spherical harmonic tomography power spectrum}

The FM for a tomographic survey employing $N$ redshift bins can be derived from Eq.~\ref{eq:FisherMat} assuming a Gaussian likelihood for the spherical harmonics coefficients. The result is \cite{2004PhRvD..70d3009H}
\begin{equation}
F_{\alpha \beta} = f_{\text{sky}} \sum_{l} \frac{\left (2 l +1 \right ) \Delta l}{2} \Tr \left [ \textbf{D}_{ l \alpha} \tilde{\textbf{C}}_{l}^{-1} \textbf{D}_{ l \beta} \tilde{\textbf{C}}_{l}^{-1} \right ]
\label{eq:FisherClz}
\end{equation}
where the sum is over bands of width $\Delta l$ in the power spectrum and we set $\Delta l = 1$. The data covariance is an $N \times N$ matrix given by 
\begin{equation}
\left [ \tilde{\textbf{C}}_{l} \right ]^{ij} = C_{l}^{x_{i} x_{j}} + N_{l}^{x_{i} x_{j}} 
\label{eq:covmat}
\end{equation}
where the $x_{i}, x_{j}$ denote the respective bins. The first term in Eq.~\ref{eq:covmat} represents the innate cosmic variance, while the second term is due to shot noise and given by $N_{l}^{x_{i} x_{j}}  = \sfrac{1}{n_{\text{A}}} \; \delta_{x_{i} x_{j}}$, where $n_{\text{A}}$ is the galaxy surface density of the survey.
 
The matrix $\textbf{D}_{ l \alpha}$ contains the dependence of the observables on the parameters $\theta_{\alpha}$ and has elements given by
\begin{equation}
\left [ \textbf{D}_{ l \alpha} \right ]^{ij} = \frac{\partial C_{l}^{x_{i} x_{j}}}{\partial \theta_{\alpha}} 
\end{equation}
The simple scaling with $f_{\text{sky}}$ accounts for the fact that angular modes become coupled for incomplete sky coverage. This reduces the number of independent modes at a given angular scale $l$ and therefore increases the uncertainties due to cosmic variance \cite{1994ApJ...421L...5S}. 

\subsection{\label{subsec:FM_Clk} The Fisher matrix for the SFB power spectrum}

The computation of the FM for the SFB power spectrum from the Gaussian likelihood for the SFB coefficients $\delta_{lm} (k)$ is challenging due to the correlations between different $k$ modes, which are due to time-evolution of the overdensity field and finite survey effects. The complication arising from the non-diagonal data covariance matrix can be dealt with in two different ways: (i) by choosing a finite grid in $k$ space and computing the FM on this discrete grid or (ii) by approximating the full FM by assuming a diagonal data covariance matrix. Drawing from previous work (\cite{1997PhRvL..79.3806T, 1997MNRAS.290..456H, 2003MNRAS.343.1327H, 2013PhRvD..88b3502Y}) we can find expressions for the FM in both cases. 

The FM for a measurement of the SFB power spectrum for $n$ discrete wave vectors $k_{i}$ can be written as \cite{2003MNRAS.343.1327H}
\begin{equation}
F_{\alpha \beta} = f_{\text{sky}} \sum_{l} \frac{\left (2 l +1 \right ) \Delta l}{2} \Tr \left [ \hat{\textbf{C}}_{l}^{-1} \frac{\partial \hat{\textbf{C}}_{l}}{\partial \theta_{\alpha}} \hat{\textbf{C}}_{l}^{-1} \frac{\partial \hat{\textbf{C}}_{l}}{\partial \theta_{\beta}} \right ]
\label{eq:Fisher_disc}
\end{equation}
where the sum is over bands of width $\Delta l$ in the power spectrum and we set $\Delta l = 1$ and the scaling with $f_{\text{sky}}$ accounts for incomplete sky coverage. $\hat{\textbf{C}}_{l}$ is the non-diagonal covariance matrix for given angular multipole $l$
\begin{equation}
\hat{\textbf{C}}_{l} =
 \begin{pmatrix}
  \tilde{C}_{l} (k_{1}, k_{1}) & \tilde{C}_{l} (k_{1}, k_{2}) & \cdots & \tilde{C}_{l} (k_{1}, k_{n}) \\
  \tilde{C}_{l} (k_{2}, k_{1}) & \tilde{C}_{l} (k_{2}, k_{2}) & \cdots & \tilde{C}_{l} (k_{2}, k_{n}) \\
  \vdots  & \vdots  & \ddots & \vdots  \\
  \tilde{C}_{l} (k_{n}, k_{1}) & \tilde{C}_{l} (k_{n}, k_{2}) & \cdots & \tilde{C}_{l} (k_{n}, k_{n})
 \end{pmatrix}
\label{eq:covmat_SFB_disc}
\end{equation}
and $\tilde{C}_{l}(k_{i}, k_{j}) = C_{l}(k_{i}, k_{j}) + N_{l}(k_{i}, k_{j})$. The first term is again the cosmic variance and the second is the shot noise in a survey with galaxy volume density $\bar{n}(r)$ given by \cite{2013PhRvD..88b3502Y}
 \begin{equation}
N_{l} (k_{i}, k_{j}) = \left (\frac{2 k_{i} k_{j}}{\pi} \right ) \int \text{d} r r^{2} \phi(r) j_{l} \left (k_{i} r \right ) j_{l} \left (k_{j} r \right ) \frac{1}{\bar{n}(r)}
\label{eq:SFB_SN}
\end{equation}
If we assume a broad window function, such that mode coupling can be neglected \cite{1994ApJ...426...23F}, we can approximate $C_{l}(k, k') = 0 $  for $k \neq k'$. This allows us to obtain a simplification of Eq.~\ref{eq:Fisher_disc} given by
\begin{equation}
F_{\alpha \beta} = f_{\text{sky}} \sum_{l} \frac{\left (2 l +1 \right ) \Delta l}{2} \int \limits_{k_{\text{min}}}^{k_{\text{max}}} \frac{L \text{d} k}{2 \pi} \frac{1}{\left ( C_{l}(k,k) + N_{l}(k,k) \right )^{2}} \frac{\partial C_{l}(k,k)}{\partial \theta_{\alpha}} \frac{\partial C_{l}(k,k)}{\partial \theta_{\beta}}
\label{eq:Fisher_cont}
\end{equation}
where the sum is over bands of width $\Delta l$ in the power spectrum, $L$ denotes the maximal length scale probed in the survey and $k_{\text{min}}$, $k_{\text{max}}$ denote the wave vector limits of the survey. For our calculations we set $L$ to the characteristic survey depth i.e. $L = r_{0}$ \footnote{The maximal length scale probed $L$ is not a well-defined quantity, but parameter constraints seem stable against changing specification, since setting $L = V^{\frac{1}{3}}_{\text{survey}}$ changes results by at most $10 \%$.} and $\Delta l = 1$.
 
For a detailed derivation of Equations \ref{eq:Fisher_disc} and \ref{eq:Fisher_cont} the reader is referred to Appendix \ref{ap:FisherMatrices}. We note that we do not include any optimal weighting of the data \cite{1995MNRAS.275..483H}, a subject which will be interesting for future work.

\section{\label{sec:Results} Results}

As a means for assessing the applicability of both the spherical harmonic tomography and the spherical Fourier Bessel power spectrum to upcoming galaxy redshift surveys, we compare their forecasted performance in a Fisher matrix analysis. From the numerous possible combinations discussed in Section \ref{sec:Baseline}, we have chosen to place our emphasis on two topics: We first focus on each statistic separately and address the main complication associated with it; then we compare the constraining power of both statistics for the baseline survey (Section \ref{sec:Baseline}).

\begin{table}
\caption{Parameter constraints obtained for different implementations of the SHT and SFB power spectrum. Results for the baseline configuration are marked in bold. Note that in the case of the SFB the full cov. results neglect contributions due to shot noise, whereas all other constraints assume shot noise as specified in Sec.~\ref{sec:Baseline}.} \label{tab:results_SHT_SFB}
\begin{center}
\begin{ruledtabular}
\begin{tabular}{cccccccccc}
Statistic & \multicolumn{2}{c}{Implementation} &Radial resolution\footnote{For the SFB power spectrum this corresponds to $k_{\text{max}}$} & $\sigma_{h}$ & $\sigma_{\Omega_{\text{m}}}$ & $\sigma_{\Omega_{\Lambda}}$ &  $\sigma_{w_{0}}$ & $\sigma_{n_{\text{s}}}$ & $\sigma_{\sigma_{8}}$ \\ \hline
                                 
\multirow{7}{*}{SHT} & Limber & $l_{\text{max}} = 50$ & 7 bins & 5.5 & 1.8 & 1.2 & 2.8 & 0.79 & 0.22 \\ 
 & no RSD & $l_{\text{max}} = 50$ & 7 bins & 2.2 & 0.32 & 0.50 & 1.2 & 1.4 & 0.15 \\ 
 & RSD & $l_{\text{max}} = 200$ & 7 bins & 0.082 & 0.040 & 0.10 & 0.23 & 0.10 & 0.030 \\ \cline{2-10}
 & \multirow{5}{*}{RSD} & $\boldsymbol{l_{\text{max}} = 50}$  & \textbf{7 bins} & \textbf{0.42} & \textbf{0.17} & \textbf{0.48} & \textbf{1.0} & \textbf{0.33} &\textbf{0.15} \\ 
  & & $l_{\text{max}} = 50$ w/ BAOs& 7 bins & 0.38 & 0.12 & 0.45 & 0.97 & 0.28 & 0.11 \\ 
 &  & $l_{\text{max}} = 50$ & 10 bins & 0.078 & 0.098 & 0.26 & 0.59 & 0.30 & 0.084 \\ 
 &  & $l_{\text{max}} = 50$ & 20 bins & 0.15 & 0.073 & 0.19 & 0.38 & 0.20 & 0.074 \\ 
 &  & $l_{\text{max}} = 50$ & 30 bins & 0.12 & 0.058 & 0.17 & 0.34 & 0.18 & 0.060 \\ \hline
\multirow{9}{*}{SFB} & \multirow{3}{*}{full cov.\footnote{Results for full Fisher matrix using full covariance matrix i.e. Eq.~\ref{eq:Fisher_disc}}} & Logarithmic & $0.2 \; h \; \text{Mpc}^{-1}$ & 0.30 & 0.048 & 0.32 & 0.46 & 0.31 & 0.67 \\ 
& & Linear & $0.2 \; h \; \text{Mpc}^{-1}$& 0.22 & 0.047 & 0.29 & 0.40 & 0.25 & 0.49 \\
& & Comoving & $0.2 \; h \; \text{Mpc}^{-1}$ & 0.38 & 0.043 & 0.55 & 1.3 & 0.38 & 0.86 \\ \cline{2-10}
& \multirow{6}{*}{diag cov.\footnote{Results for continuous Fisher matrix using diagonal covariance matrix i.e. Eq.~\ref{eq:Fisher_cont}}} & Logarithmic & $0.2 \; h \; \text{Mpc}^{-1}$ & 0.33 & 0.059 & 0.52 & 0.71 & 0.30 & 0.74  \\ 
& & Linear & $0.2 \; h \; \text{Mpc}^{-1}$ & 0.53 & 0.086 & 0.82 & 1.0 & 0.39 & 1.2  \\ 
& & \textbf{Comoving} & $\mathbf{0.2} \; \boldsymbol{h} \; \textbf{Mpc}^{-1}$ & \textbf{0.37} & \textbf{0.091} & \textbf{0.72} & \textbf{2.7} & \textbf{0.32} & \textbf{0.81}\\
& & Comoving w/ BAOs& $0.2 \; h \; \text{Mpc}^{-1}$ & 0.11 & 0.028 & 0.31 & 1.2 & 0.15 & 0.26 \\ 
& & Comoving & $0.15 \; h \; \text{Mpc}^{-1}$ & 0.48 & 0.097 & 0.79 & 2.8 & 0.51 & 1.1 \\
& & Comoving & $0.1 \; h \; \text{Mpc}^{-1}$ & 0.55 & 0.10 & 0.92 & 2.9 & 0.62 & 1.2 \\
\end{tabular}
\end{ruledtabular}

\end{center}
\end{table} 

\subsection{\label{subsec:Res_Clz} Spherical harmonic tomography power spectrum}

The constraints on cosmological parameters obtained when analysing the baseline survey (\ref{sec:Baseline}) using the SHT power spectrum and only taking into account auto-power spectra \footnote{We find that including the cross-correlations does not affect results significantly and we thus neglect them in order to match the specifications used for the SFB power spectrum more closely.} are highlighted in Table \ref{tab:results_SHT_SFB}. For the baseline configuration we obtain uncertainties of the order of the parameter value, which is due to the fact that we only consider large-scale information from angular multipoles $l \in [2, 50]$ \footnote{We believe that this reduced range does not affect our results because we are mainly concerned with comparing two different statistics.}. Increasing the maximal angular scale probed to $l_{\text{max}} = 200$, which corresponds to the non-linear cutoff for the lowest redshift bin, considerably improves parameter constraints. The restriction to large-scale angular perturbations is due to the calculation of the SFB power spectrum, which becomes slow for smaller scales. Nonetheless, the matter density of the universe and the power spectrum amplitude are already sensibly constrained whereas the dark energy sector is poorly constrained due to the significant degeneracies present. Before comparing these results to the constraints obtained for the SFB power spectrum, we first discuss the main complication associated with tomographic analyses.

The SHT power spectrum necessitates tomographic analyses of galaxy catalogues, which amounts to splitting the data into redshift bins. This additional freedom raises the question of how to optimally perform this subdivision. For a fixed baseline survey and therefore data, we expect to see small differences between binning schemes. As we will show below, on the contrary, we can identify instabilities when implementing different bin configurations for a given survey, when we estimate their respective constraining power in a Fisher analysis using Eq.~\ref{eq:FisherClz}. Not only is this behaviour unexpected but it also implies that these instabilities need to be kept in mind when e.g. comparing the forecasted performance of future surveys.

As an example for studying the effects of redshift binning on parameter constraints, we investigate configurations that differ in the amount of redshift bin overlap. In general there are two causes for bin overlap in galaxy redshift surveys: (i) in spectroscopic surveys, bins can purposely be tailored to have overlap while (ii) in photometric surveys, redshift bins will overlap due to inaccurate redshift measurements. We focus on spectroscopic surveys and therefore only consider case (i). A nonzero overlap between redshift bins will cause them to be correlated, if we assume that they are both located in the same part of the sky. Investigating the impact of bin overlap/correlation on the constraining power of galaxy redshift surveys therefore not only highlights instabilities with data binning but it also addresses the core of the same sky-different sky issue (see e.g. \cite{2013arXiv1307.8062K, 2013arXiv1308.6070D}), which is the question of how much correlations between data sets can affect parameter constraints. 

We investigate the effects of bin configuration on constraining power using a series of highly simplified toy models, which are based on the Limber approximation and ignore shot noise contributions. For a detailed description of these, the reader is referred to Appendix \ref{ap:TM}. As shown in \ref{ap:TMI}, we find that increasing the amount of overlap between bins, while keeping their mean redshifts fixed and taking into account correlations, can result in an improvement of cosmological parameter constraints by as much as a factor of 2. This behaviour is only found when constraining parameters that exhibit a high level of redshift degeneracy between each other i.e. parameters which can only be distinguished with information at separate redshifts; an example from cosmology is the redshift degeneracy between parameters which control the growth of structure as a function of time and the overall clustering amplitude. On the other hand, we find that constraints on non-redshift degenerate parameters as well as fixed errors are insensitive to changes in bin overlap. It is important to point out that these conclusions do not apply to bin overlap caused by redshift uncertainties (case (ii)). Redshift errors cannot be modelled solely as a redshift bin broadening, since this approach does not take into account the uncertainty introduced in the redshift distribution. If redshift uncertainties are implemented as in \cite{2007MNRAS.381.1018A}, we find that increased bin overlap due to larger redshift uncertainties worsens parameter constraints as intuitively expected. 

These results suggest that the main effect of overlap between redshift bins on spectroscopic surveys is to break redshift degeneracies between parameters. This seems counterintuitive but as we show in Appendix \ref{ap:TMII}, the dependence of parameter constraints on correlation is a generic feature of such data sets. This suggests that the observed sensitivity of parameter constraints on binning scheme is due to the fact that the amount of correlation between redshift bins, which has an effect on parameter constraints, is scheme dependent. 

The results presented so far have been based on tomographic analyses consisting of only two redshift bins; as the number of redshift bins is increased, the effect of bin overlap becomes negligible as shown in Fig.~\ref{fig:TM1}. The more available cosmological information is recovered from a survey, the less sensitive parameter constraints become to binning schemes. In order to obtain stable parameter constraints from a tomographic analysis of galaxy redshift surveys it is therefore essential to ensure that the available information is well recovered by the survey. We will review the limitations imposed on a tomographic analysis returning back to our baseline survey.

\begin{figure}[h]
\begin{center}
\includegraphics[scale=0.5]{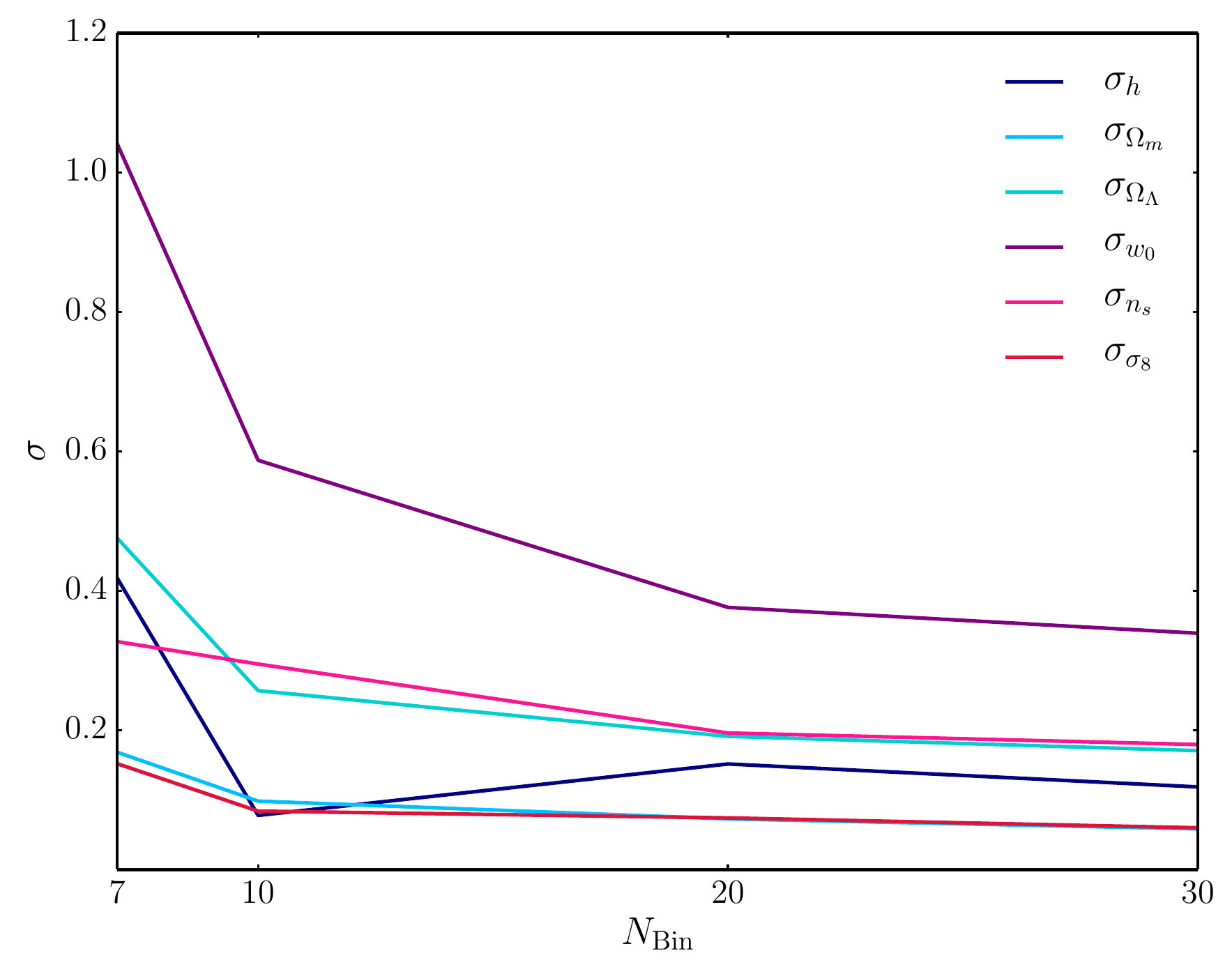}
\caption{\label{fig:FMs_bins} Uncertainties ($1 \; \sigma$) on cosmological parameters obtained with SHT power spectrum as a function of the number of redshift bins $N_{\text{Bin}}$.}
\end{center}
\end{figure}

The parameter that mainly controls the constraining power of a tomographic survey is its radial resolution; in practice this is the number of redshift bins. Starting from the baseline survey, we increase the number of redshift bins from $N_{\text{Bin}} = 7$ to $N_{\text{Bin}} = 30$ as shown in Table \ref{tab:results_SHT_SFB} and Fig.~\ref{fig:FMs_bins}. This reduces uncertainties by almost a factor of three, which is in agreement with the rough $\sfrac{1}{\sqrt{N_{\text{Bin}}}}$ scaling of parameter constraints with bin number when shot noise is not yet dominant \cite{2014JCAP...01..042D}. Therefore the maximal cosmological information retrievable analysing a survey using the SHT power spectrum is limited by redshift accuracy and shot noise.

As seen from Table \ref{tab:results_SHT_SFB} we reanalyse the baseline survey neglecting redshift space distortions; once performing the full calculation and once assuming the Limber approximation. The results indicate that both changes deteriorate parameter constraints, showing that including RSDs in tomographic analyses increases the amount of cosmological information. Furthermore, the results obtained with the Limber approximation and the exact calculation deviate significantly (differences of up to a factor of 5), which suggests that approximations in power spectrum calculations should be used carefully when computing Fisher matrices. 

\subsection{\label{subsec:Res_SFB} Spherical Fourier Bessel power spectrum}

The constraints on cosmological parameters obtained when analysing the baseline survey (\ref{sec:Baseline}) using the spherical Fourier Bessel power spectrum are emphasised in Table \ref{tab:results_SHT_SFB}. In agreement with the results for the SHT power spectrum, we obtain constraints of the same order of magnitude as the parameters themselves, an effect which we again attribute to the small multipole range considered. The best-constrained parameter is the matter density $\Omega_{\text{m}}$, whereas the SFB analysis mostly yields larger uncertainties on the remaining cosmological parameters than its tomographic counterpart. As in the previous section, we address particular issues associated with this analysis before turning to the comparison between the two methods.

\subsubsection{\label{subsubsec:FMComparisons} Comparison of Fisher matrix computation techniques}

The computation of Fisher matrices for the SFB power spectrum through Eq.~\ref{eq:Fisher_disc} \footnote{There is one subtlety involved with Fisher matrix calculations for the SFB power spectrum: as can be seen from Eq.~\ref{eq:SFB_SN}, the shot noise contribution to the power spectrum is essentially a galaxy number count and therefore cosmology dependent. Since we ignore information from non-linear scales, we won't include this information when estimating the constraining power of a survey.} is time-consuming and numerically challenging because it requires inverting the covariance matrix defined in Eq.~\ref{eq:covmat_SFB_disc}. This step is tricky for two reasons: Firstly, as seen from Fig.~\ref{fig:Cls_SFB}, the SFB power spectrum falls off sharply for large scales with $k \ll \sfrac{l}{r_{0}}$ for given angular scale $l$ and survey depth $r_{0}$. This is because in a cone of angular extent $\theta \sim \sfrac{1}{l}$, large radial modes cannot be measured when the survey depth is finite. The rapid decrease in large scale power results in a considerable dynamic range in the covariance matrix eigenvalues, making it almost singular. Secondly, neighbouring wave vectors $k$ are strongly correlated, which further complicates the inversion of the covariance matrices. 

In order to estimate the SFB Fisher matrix through Eq.~\ref{eq:Fisher_disc} it is therefore inevitable to restrict calculations to separated wave vectors with $k \geq \sfrac{l}{r_{0}}$ to overcome numerical instabilities. In practice we cut off all large-scale information for each angular scale $l$ as soon as it causes the covariance matrix condition number, which is a measure for accuracy loss in matrix inversion, to exceed $\kappa_{\text{crit}} = 10^{2}$. This is possible but it seems desirable to investigate alternatives to this ``brute-force'' approach. 

Even though neighbouring wave vectors are strongly correlated, the correlations tend to rapidly decrease as we move away from the diagonal. This suggests resorting to the approximation $C_{l} \left (k, k' \right ) \simeq 0$ for $k \neq k'$ in order to obtain useful approximations to Eq.~\ref{eq:Fisher_disc}. The most straight-forward implementation of these ideas is given in Eq.~\ref{eq:Fisher_cont}. Despite being an approximation to the full Fisher matrix given in Eq.~\ref{eq:Fisher_disc}, we find that Eq.~\ref{eq:Fisher_cont} yields no-shot noise as well as shot noise constraints which are mostly accurate to better than a factor of 2 for the baseline survey (for detailed results, see Table \ref{tab:full_vs_approx} in Appendix \ref{ap:results_FMs}). These results agree with those obtained for the Cartesian matter power spectrum \cite{1997PhRvL..79.3806T} and encourage the use of Eq.~\ref{eq:Fisher_cont} for fast calculations which allow errors of up to a factor of a few. 

\begin{figure}
\includegraphics[scale=0.5]{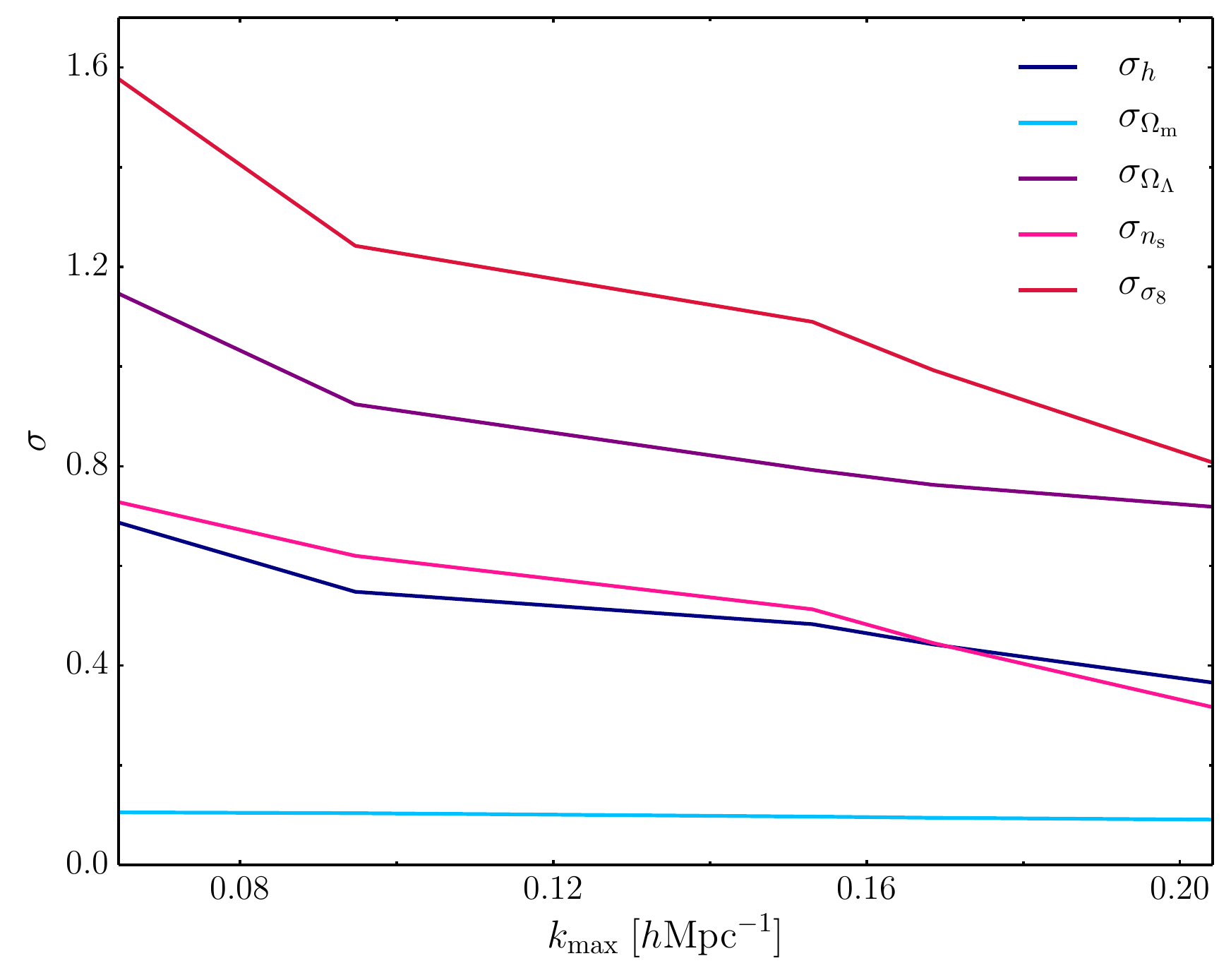}
\caption{\label{fig:FMs_SFB} Uncertainties ($1 \; \sigma$) on cosmological parameters obtained using the SFB power spectrum as a function of maximal wave number $k_{\text{max}}$. Constraints on $w_{0}$ are not shown for clarity.}
\end{figure}

Making use of this simplifying approximation, we investigate the impact of the non-linearity wave-vector cut on parameter constraints. As can be seen from Table \ref{tab:results_SHT_SFB} and Fig.~\ref{fig:FMs_SFB} decreasing the maximal wave vector $k_{\text{max}}$ by a factor of two increases parameter uncertainties by almost the same amount. The increase is larger than theoretically expected, since decreasing the cutoff scale by a factor of two will halve the number of available modes and should thus lead to an increase in uncertainty by a factor $\sqrt{2}$. We believe that this is due to the fact that increasing the cutoff scale additionally helps breaking parameter degeneracies, since we approximately observe the theoretical scaling for the fixed parameter constraints.

\subsubsection{\label{subsubsec:Distance_measure} The choice of distance-redshift relation}

As discussed in Sec.~\ref{subsubsec:Alt_Distances}, the need for assuming a fiducial cosmological model can be mitigated by analysing surveys using distance-redshift relations that can directly be computed from observable quantities. Using the expressions derived in Sec.~\ref{subsubsec:Alt_Distances} we can investigate the impact of the distance-redshift relation on the obtained power spectra as well as survey constraining power.

We focus on two simple alternatives to the comoving distance as given in Eq.~\ref{eq:SFB_Distance_meas}. Both approximations are fairly accurate at low redshift; at very high redshifts on the other hand, both approximations break down because they considerably overestimate the comoving separation. The SFB power spectra obtained with these two distance-redshift relations are shown in Fig.~\ref{fig:Cls_SFB_dist}. The choice of a different fiducial distance causes a shift in the observed SFB power spectra, because the window functions are offset from those in comoving distance.

These simple distance-redshift relations are viable alternatives to the comoving distance only if analysing a survey in terms of them does not significantly reduce its constraining power. To test their performance, we compare their constraints on cosmological parameters for a survey as defined in \ref{sec:Baseline} in two different ways: Since the continuous Fisher matrix is an acceptable approximation to the full calculation, we will employ it to compare the forecasted parameter constraints obtained with all three distance measures taking shot noise into account. As a mere illustration, we additionally compare the constraints obtained from the full Fisher matrix, neglecting any shot noise contributions.

Both these results are shown in Table \ref{tab:results_SHT_SFB}. The constraints are similar, irrespective of the distance-redshift relation chosen. The only parameter exhibiting a significant dependence on the way distance is related to redshift is the dark energy equation of state parameter $w_{0}$. An analogous behaviour is seen in the no shot-noise constraints obtained with the full Fisher matrix: we obtain different constraints especially for those parameters, which the comoving distance depends on, whereas the remaining constraints are largely insensitive to the distance-redshift relation of choice. Choosing a distance-redshift relation other than the comoving distance therefore appears to cause the SFB power spectrum to exhibit a stronger dependence on these former parameters because the volume element and the distance in Eq.~\ref{eq:SFB_window_noRSD} and \ref{eq:SFB_window_RSD} do not change in the same way.

As we include contributions due to shot noise, this potential constraining power is considerably reduced because alternative distance-redshift relations tend to overestimate the comoving separations for large redshift. The shot noise is therefore enhanced, which largely removes the gain from increased sensitivity.

The above considerations illustrate that constraints obtained from an SFB analysis of a galaxy redshift survey seem to be mostly stable against changes in the assumed distance-redshift relation. This suggests that it could be possible to analyse galaxy clustering using distance-redshift relations which only depend on observable quantities, without too large a loss in constraining power.

\subsection{\label{subsec:comparisons} Comparison between spherical harmonic tomography and SFB power spectrum}

After focusing on each of the two statistics separately we can compare their baseline constraints shown in Table \ref{tab:results_SHT_SFB}. Unexpectedly, the survey constraining power is significantly affected by the choice of analysis method: When baryon acoustic oscillations are neglected, the SFB power spectrum yields weaker constraints, particularly on those cosmological parameters that need redshift leverage in order to be distinguished (i.e. growth and amplitude parameters). This is already evident for our baseline but the effect is enhanced if we consider larger bin numbers in the SHT analysis.

The weakness of constraints on growth as well as amplitude of matter fluctuations seems to be a generic feature of SFB analyses: When the underlying field is statistically isotropic and homogeneous (SIH), the SFB coefficients are given by Eq.~\ref{eq:SFB_Cart} i.e. they correspond to an angular average of the Cartesian Fourier coefficient. In practice, the SIH condition is not met due to time evolution of the overdensity field and finite survey effects and the SFB coefficients are related to their Cartesian counterparts through Eq.~\ref{eq:SFBcoeffs_noSIH}. Any violation of the SIH condition will therefore introduce a coupling between the considered radial and angular scales $k$, $l$ and the redshift at which $\delta \left( \mathbf{k} \right )$ will mostly be measured. Around $r_{0}$, the decrease in the selection function amplitude breaks the SIH condition which causes modes to add incoherently and leads to cancellations. The contributions to the SFB power spectrum will therefore be preferentially weighted toward lower redshift, which results in smaller redshift leverage and weaker combined constraints.

These results therefore suggest that, even though both analysis methods are equivalent for infinite survey extent and recovery of all available modes, because the information content of the overdensity field does not depend on the basis set in which it is analysed, they appear not to be equivalent for finite surveys. Analyses of galaxy redshift surveys through the SFB power spectrum are more affected by finite survey effects, which means that some high redshift information will be downweighted. The SHT power spectrum analysis, on the other hand, allows us to probe more efficiently the complete high redshift range of the survey.

We test the impact of baryon oscillations on our results by reanalysing our baseline configurations using the transfer function as specified in \cite{Eisenstein:1998aa}. From Table \ref{tab:results_SHT_SFB} we see that adding BAOs improves the SFB constraints, while leaving the SHT power spectrum constraints mostly unchanged. This suggests that a significant fraction of the information lost due to finite survey effects in SFB analyses can be compensated by the fact that its 3 dimensional nature allows us to recover the information contained in the BAOs, while an SHT analysis tends to dilute these features \footnote{The change of transfer function from \cite{1995ApJS..100..281S} to \cite{Eisenstein:1998aa} without BAOs (the so-called ``no-wiggles'' approximation obtained with fractional baryon density $\Omega_{\text{b}} \neq 0 $ but ignoring the oscillatory contribution to the power spectrum) does not significantly affect parameter constraints, which suggests that the improvement in constraining power can be mainly attributed to the presence of BAOs.}. The inclusion of BAOs results in comparable constraints for both methods. This suggests that SFB is better suited for capturing information beyond the smooth shape of the power spectrum.

\subsection{\label{subsec:implementation} Implementation effects on estimated survey constraining power}

The constraining power of a particular survey, as estimated from Fisher matrix calculations, is clearly determined by survey specifications and included physics. Nevertheless, details in the particular implementation can also affect parameter constraints and we can investigate the magnitude of this effect using our simplified toy model (see Appendix \ref{ap:TMI}). Our findings suggest that the magnitude can potentially equal that of changing survey specifications, when one of the following two conditions is fulfilled: (i) As illustrated in Appendix \ref{ap:TMI}, the choice of binning scheme and thus implementation ceases to be relevant as more information is retrieved from a particular survey and parameter constraints become tighter. This therefore suggests that the choice of prior can significantly affect the stability of parameter constraints: applying a tight prior reduces the susceptibility of parameter constraints to implementation. (ii) A second essential choice is the set of constrained parameters: as highlighted by the toy model, parameters which are redshift degenerate with each other are particularly affected by changes in implementation. On the other hand, constraints on non-degenerate parameters are expected to be more stable. It is thus important to be aware of these instabilities whenever constraining a set of degenerate and loosely constrained parameters. 

These susceptibilities can further be amplified due to numerical instabilities in Fisher matrix calculations. Fisher matrices can have large condition numbers $\kappa \left (F \right )$ i.e. be close to numerically singular, if one or more parameters are not well constrained by the data. Therefore the error introduced by the Fisher matrix inversion can be of order $100 \%$, if the accuracy in the Fisher matrix elements is smaller than $\kappa \left (F \right )^{-1}$ \cite{2008PhRvD..77d2001V}.

These findings further suggest that care has to be taken when comparing Fisher matrix results. Whenever two different results need to be compared it is essential to make sure that not only the survey specifications are similar but also that priors and set of constrained parameters agree with each other. This is relevant in light of the recent discussion regarding benefits of performing spectroscopic and photometric galaxy surveys in the same part of the sky (see e.g. \cite{2013arXiv1307.8062K, 2013arXiv1308.6070D}): a reliable comparison between the results obtained by different groups seems difficult due to the differing choices of priors and constrained parameters. Exactly matching the baseline survey is especially important in this case since it investigates the impact of cross-correlations on parameter constraints, which, as indicated by the toy model, only has an effect when parameters are loosely constrained and degenerate with each other. This suggests that changes in implementation have the potential to even affect qualitative results in this particular case.

\section{\label{sec:Conclusions} Conclusions}

Using a Fisher analysis, we have investigated 3-dimensional analyses of galaxy redshift surveys on a spherical sky. In the course of our analysis, it has become evident that Fisher matrix results need to be carefully analysed and compared. Especially when Fisher matrix methods are used to forecast constraints on large parameter sets, exhibiting degeneracies among one another, the obtained constraints are susceptible to details in implementation.

In particular we have compared the SHT and the SFB power spectrum, two statistics that are designed for the analysis of galaxy redshift surveys in a spherical geometry. By comparing their forecasted constraints on cosmological parameters, we have shown the applicability of approximations, such as the Limber approximation, and the numerical issues associated with computing these statistics. We have also studied the sensitivity of these statistics to the detail of their implementation. Our analysis is based on several simplifying assumptions; in particular, we restrict ourselves to linear, unbiased galaxy clustering and only focus on large scale power spectrum modes. For future work it would be interesting to include a treatment of these effects.
 
Using toy models, we find that constraints obtained from a tomographic analysis of galaxy redshift surveys can be susceptible to implementation effects like redshift bin overlap, if only a limited amount of the total available information is retrieved. This suggests that in order to be stable against changes in implementation, it is important to retrieve as much information as possible from the tomographic analyses, e.g. by using a large number of redshift bins. 

The computation of the SFB power spectrum from data relies on the assumption of a distance-redshift relation, usually the cosmology-dependent comoving distance. Analyses of galaxy redshift surveys by means of the SFB power spectrum therefore require the assumption of a cosmological model, prior to testing it. Comparing the SFB parameter constraints obtained using alternative distance-redshift relations, we find them to be largely stable against changes in the assumed distance. This suggests that future surveys could in principle be analysed using distance-redshift relations only relying on observable quantities without too large a degradation in parameter constraints.

For the baseline survey configuration we considered, we find that the SHT power spectrum yields somewhat tighter constraints than the SFB power spectrum. When we add baryon oscillations, on the other hand, the two methods yield comparable constraints. We attribute the former to the fact that the SFB power spectrum is less sensitive to modes at high redshift near the survey boundary, while the SHT power spectrum can be taylored to probe these modes. In the presence of BAOs, this effect can be compensated by the fact that the 3 dimensional nature of the SFB transform allows us to resolve the baryonic oscillations, which tend to be diluted in SHT analyses. This fact would make SHT analyses advantageous for future spectroscopic galaxy redshift surveys mainly focusing on the power spectrum shape, while the SFB power spectrum may be well suited for specific applications like BAOs.

\begin{acknowledgments}
The authors would like to thank Jo\"{e}l Ackeret, Enea Di Dio, Lukas Gamper, Francesco Montanari, Ana\"{i}s Rassat and Jaiyul Yoo for useful discussions and an anonymous referee for comments and suggestions that have improved this paper. This work was supported in part by grant $200021\_143906$ from the Swiss National Science Foundation.
\end{acknowledgments}

\appendix

\section{\label{ap:Radialisation_SFB} The radialisation of the SFB power spectrum in the presence of RSDs}

The overdensity field in the absence of RSDs at a constant time $r$ is isotropic and homogeneous and can be directly related to the Cartesian matter power spectrum as in Eq.~\ref {eq:SFB_PS} \cite{2005PhRvD..72b3516C}. The isotropy is broken in the presence of RSDs but an approximate relation between these two quantities still holds. The SFB coefficients of the overdensity field are related to their Cartesian Fourier counterpart through
\begin{equation}
\delta_{lm} (k, r) =  \frac{1}{\sqrt{8 \pi ^{3}}} k i^{l} \int \text{d} \Omega_{k} \delta \left( \mathbf{k},  r\right ) Y^{*}_{lm}(\theta_{k}, \phi_{k} ) 
\label{eq:SFB_Cart}
\end{equation}
The contribution to the Cartesian Fourier coefficient due to RSDs is given by
\begin{equation}
\delta_{\text{RSD}} \left( \mathbf{k},  r\right ) = \beta \mu_{k}^{2} \delta \left( \mathbf{k},  r\right )
\end{equation}
with a power spectrum $\langle \delta_{\text{RSD}} \left( \mathbf{k},  r\right ) \delta_{\text{RSD}}^{*} \left( \mathbf{k'},  r\right ) \rangle = \beta^{2} \mu_{k}^{4} P(k, r) \: \Dirac(\mathbf{k'}-\mathbf{k})$. The quantity $\mu_{k}$ is the cosine of the angle between the line of sight direction and the wave vector direction. In the flat-sky limit we can assume that the line of sight direction is constant. For a fixed angular multipole $l$, the SFB power spectrum will obtain contributions from increasingly radial wave vectors for larger $k$. In the flat-sky limit and large radial wavevectors $k$ we can thus approximate $\mu_{k} \simeq 1$. Inserting this into Eq.~\ref{eq:SFB_Cart} using the identities
\begin{equation}
\Dirac(\mathbf{k'}-\mathbf{k}) = \frac{1}{(2 \pi)^{3}} \int \text{d}^{3} \textbf{x} e^{(\mathbf{k'}-\mathbf{k}) \cdot \mathbf{x}} 
\end{equation}
and \cite{Abramowitz1964} 
\begin{equation}
e^{i \textbf{k} \cdot \textbf{r}} = 4 \pi \sum_{l, m} i^{l} j_{l}(k r) Y_{lm}^{*} (\hat{\textbf{k}} ) Y_{lm}\left (\hat{\textbf{n}} \right )
\label{eq:Exponential}
\end{equation}
gives the contribution to the SFB power spectrum due to RSDs
\begin{equation}
\langle \delta_{lm} (k, r) \delta^{*}_{l'm'} (k', r) \rangle = \beta^{2} P(k, r) \: \Dirac(k'-k) \delta_{ll'} \delta_{mm'}
\end{equation}
In the flat-sky and large wave vector limit even the RSD contribution approximately radialises in absence of a selection function and time-dependence of the overdensity field. This behaviour is perceivable in Fig.~4 of \cite{2013PhRvD..88b3502Y}, illustrating that the curvature of the sky is negligible for small scale perturbations.

\section{\label{ap:Alt_Distances} The SFB power spectrum for a generic distance-redshift relation}

The need for assuming a cosmological model before testing it can be avoided choosing a distance-redshift relation which does not depend on cosmology.
In order to derive an expression for the SFB power spectrum, we assume a generic distance-redshift relation defined as $\tilde r (z)$ where $\tilde r$ is an arbitrary monotonic function of the redshift $z$. The measured redshift will be affected by peculiar galaxy velocities $\textbf{v}$ along the line of sight $\textbf{n}$ \cite{peacock:1999} i.e.
\begin{equation}
z_{\text{obs}} \simeq z_{\text{true}} + \frac{\textbf{v} \cdot \textbf{n}}{a c}
\end{equation}
where $c$ is the speed of light. The distance $s$ inferred from the galaxy redshifts therefore becomes
\begin{equation}
s = \tilde r (z_{\text{obs}}) \simeq \tilde r(z_{\text{true}}) +  \frac{\textbf{v} \cdot \textbf{n}}{a c} \frac{\text{d} \tilde r}{\text{d} z}
\end{equation}
The overdensity field can be decomposed in the SFB basis set with coefficients given by
\begin{equation}
\delta_{lm} (\nu) = \sqrt{\frac{2}{\pi}} \int \text{d}^{3} \textbf{s} \phi '(s) \delta ( \textbf{s} ) \nu j_{l} \left (\nu s \right ) Y^{*}_{lm}(\theta, \phi ) 
\label{eq:SFBcoeffs1}
\end{equation}
where $\nu$ denotes the wave vector conjugate to $\tilde r$ and $\phi'$ is the selection function in the new coordinate system. Since the overdensity field is independent of the distance measure we have $\text{d}^{3} \textbf{s} \: \delta \left (\textbf{s} \right ) = \text{d}^{3} \textbf{r} \: \delta \left (\textbf{r} \right )$ where $r$ is the comoving distance. Eq.~\ref{eq:SFBcoeffs1} therefore reduces to
\begin{equation}
\delta_{lm} (\nu) = \sqrt{\frac{2}{\pi}} \int \text{d}^{3} \textbf{r} \phi'(s) \delta ( \textbf{r} ) \nu j_{l} \left (\nu s \right ) Y^{*}_{lm}(\theta, \phi ) 
\label{eq:SFBcoeffs2}
\end{equation}
Following \cite{1997MNRAS.290..456H}, the functions of $s$ can be expanded as
\begin{eqnarray}
&& \phi'(s) \simeq \phi'(\tilde r) + \frac{\text{d} \phi'}{\text{d} \tilde r} \left ( \frac{\textbf{v} \cdot \textbf{n}}{a c}  \frac{\text{d} \tilde r}{\text{d} z} \right ) \label{eq:SFB_Taylor} \\
&& j_{l} \left (\nu s \right ) \simeq j_{l} \left (\nu \tilde r \right ) + \frac{\text{d} j_{l} \left (\nu \tilde r \right )}{\text{d} \tilde r} \left ( \frac{\textbf{v} \cdot \textbf{n}}{a c} \frac{\text{d} \tilde r}{\text{d} z}  \right ) \nonumber 
\end{eqnarray}
which can be inserted into Eq.~\ref{eq:SFBcoeffs2} to yield to first order
\begin{align}
\delta_{lm} (\nu) &= \sqrt{\frac{2}{\pi}} \frac{1}{(2 \pi)^{3}} \left \{ \int \text{d}^{3} \textbf{r} \int \text{d}^{3} \textbf{k}' \phi'(\tilde r) \delta ( \textbf{k}' ) e^{i \textbf{k}' \cdot \textbf{r}} \nu j_{l} \left (\nu \tilde r \right ) Y^{*}_{lm}(\theta, \phi ) \right. \nonumber \\
& + \left. \int \text{d}^{3} \textbf{r} \int \text{d}^{3} \textbf{k}' \frac{\textbf{v} ( \textbf{k}' ) \cdot \textbf{n}}{a c} e^{i \textbf{k}' \cdot \textbf{r}} \nu \frac{\text{d} \tilde r}{\text{d} z} \frac{\text{d}}{\text{d} \tilde r} [\phi'(\tilde r) j_{l} \left (\nu \tilde r \right ) ] Y^{*}_{lm}(\theta, \phi ) \right \}
\end{align}
The linear continuity equation allows us to relate the Fourier transform of the galaxy velocity field to the overdensity through \cite{dodelson:2003}
\begin{equation}
\textbf{v} (\textbf{k}) = i \beta \frac{a H(z) \delta(\textbf{k})}{k^{2}} \textbf{k}
\label{eq:LinearVelocity}
\end{equation}
Thus the SFB power spectrum for the distance-redshift relation $\tilde r (z)$ reduces to
\begin{equation}
C_{l} (\nu, \nu') = \left (\frac{2}{\pi} \right )^{2} \int \text{d} k'' k''^{2} P_{0}(k'') \left ( W_{l}(\nu, k'') + W^{r}_{l}(\nu, k'') \right ) \left ( W_{l}(\nu', k'') + W^{r}_{l}(\nu', k'') \right )
\label{eq:SFB_dist}
\end{equation}
$W_{l}(\nu, k'')$ is the real-space window function whereas $W^{r}_{l}(\nu, k'')$ accounts for the corrections due to RSDs; they are given by
\begin{equation}
W_{l} (\nu, k'') =  \int \text{d} r r^{2} D(r) \phi'(\tilde r) \nu j_{l} \left (\nu \tilde r \right ) j_{l} \left (k'' r \right )
\end{equation}
\begin{align}
W^{r}_{l} (\nu, k'') &=  \int \text{d} r r^{2} \beta \frac{\nu^{2}}{k''} D(r) \phi'(\tilde r) \frac{\text{d} \tilde r}{\text{d} z} \frac{H(z)}{c} \left [ \frac{l^{2}}{(2l+1)^{2}} j_{l-1} \left (\nu \tilde r \right ) j_{l-1} \left (k'' r \right ) \right. \nonumber \\ 
& - \left. \frac{l(l+1)}{(2l+1)^{2}} \left \{ j_{l-1} \left (\nu \tilde r \right ) j_{l+1} \left (k''r \right ) + j_{l+1} \left (\nu \tilde r \right ) j_{l-1} \left (k''r \right ) \right \} + \frac{(l+1)^{2}}{(2l+1)^{2}} j_{l+1} \left (\nu \tilde r \right ) j_{l+1} \left (k''r \right ) \right ]  \\ 
& + \int \text{d}r r^{2} \beta \frac{\nu}{k''} D(r) \frac{\text{d} \phi'}{\text{d} \tilde r} \frac{\text{d} \tilde r}{\text{d} z} \frac{H(z)}{c} \left ( \frac{l}{(2l+1)}  j_{l} \left (\nu \tilde r \right ) j_{l-1} \left (k''r \right ) - \frac{(l+1)}{(2l+1)}  j_{l} \left (\nu \tilde r \right ) j_{l+1} \left (k'' r \right ) \right ) \nonumber 
\end{align}
Since the selection function transforms as a scalar, $\phi' \left ( \tilde r\right )$ is related to the selection function in comoving distance through $\phi' \left ( \tilde r(z) \right ) = \phi \left (r(z) \right )$. 

Any measurement of the SFB power spectrum will be affected by shot noise. The number of galaxies $N_{\text{gal}}$ observed in a given survey is independent of the distance-redshift relation of choice i.e.
\begin{equation}
N_{\text{gal}} = \int \text{d}^{3} \tilde{\textbf{r}} \;  \phi' \left ( \tilde r \right ) n' \left( \tilde r \right ) = \int \text{d}^{3} \textbf{r} \; \phi \left ( r \right ) n \left( r \right ) 
\label{eq:galaxy_no}
\end{equation}
where $n \left( r \right )$ is the galaxy volume density in comoving coordinates and $n' \left( \tilde r \right )$ is the volume density in terms of $\tilde r$. Together with the identity $\phi' \left ( \tilde r(z) \right ) = \phi \left (r(z) \right )$, this implies
\begin{equation}
n' \left( \tilde r \right ) = \left \vert \frac{\text{d}^{3} r}{\text{d}^{3} \tilde r} \right \vert n \left( r \right ) 
\label{eq:galaxy_density}
\end{equation}
For a generic distance-redshift relation the shot-noise therefore reduces to
\begin{equation}
N_{l} \left( \nu, \nu' \right ) = \frac{2 \nu \nu'}{\pi} \int \text{d} r \; r^{2} \phi' \left ( \tilde r \right ) \frac{1}{n' \left( \tilde r \right )} j_{l} \left (\nu \tilde r \right ) j_{l} \left (\nu' \tilde r \right ) =  \frac{2 \nu \nu'}{\pi} \int \text{d} r \; \tilde r^{2} \left \vert \frac{\text{d} \tilde r}{\text{d} r} \right \vert \phi \left ( r \right ) \frac{1}{n \left( r \right )} j_{l} \left (\nu \tilde r \right ) j_{l} \left (\nu' \tilde r \right ) 
\end{equation}

\begin{figure}
\begin{center}
\subfigure[Logarithmic distance-redshift relation]{\includegraphics[width=0.49\textwidth]{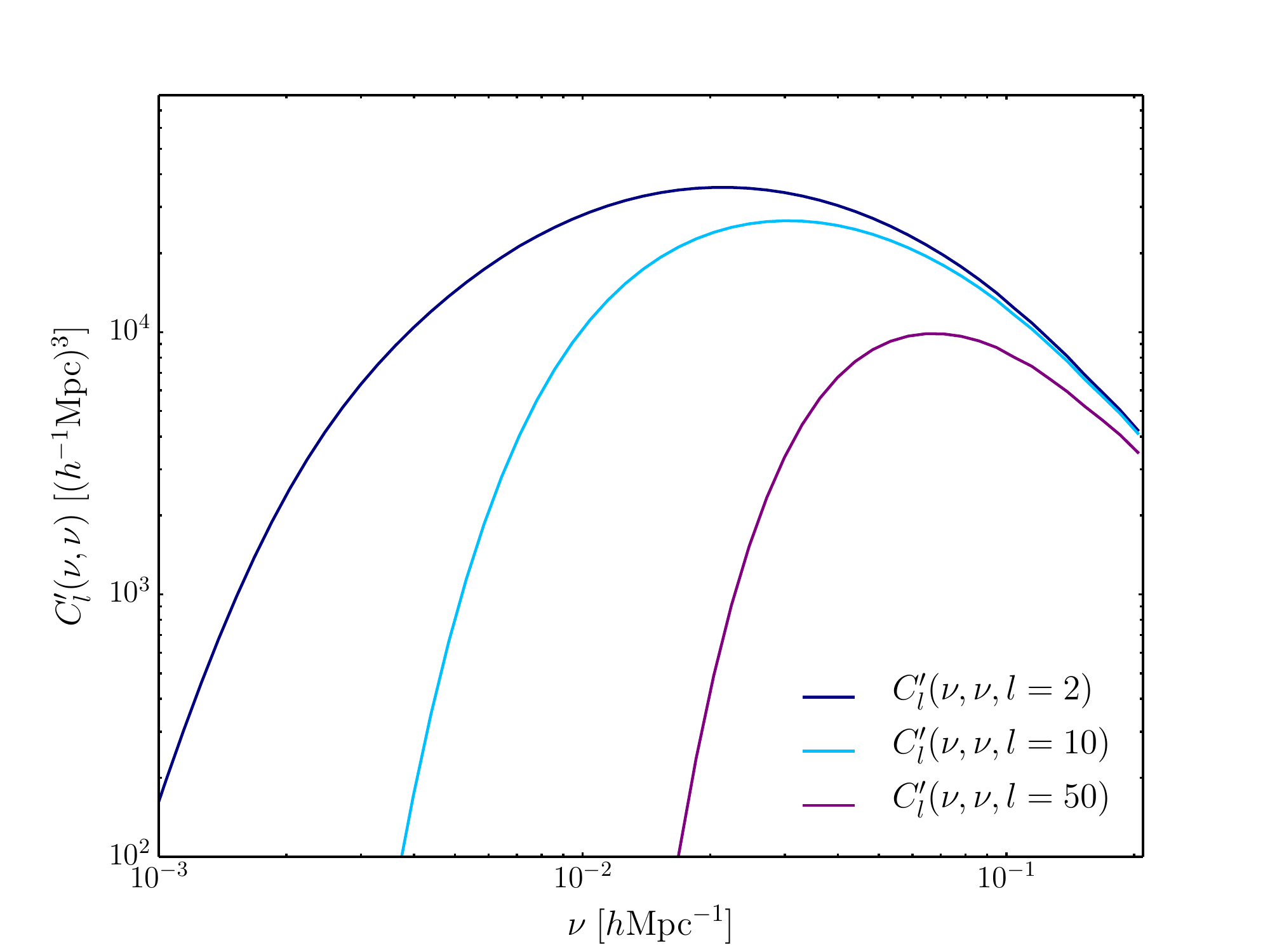}}
\subfigure[Linear distance-redshift relation]{\includegraphics[width=0.49\textwidth]{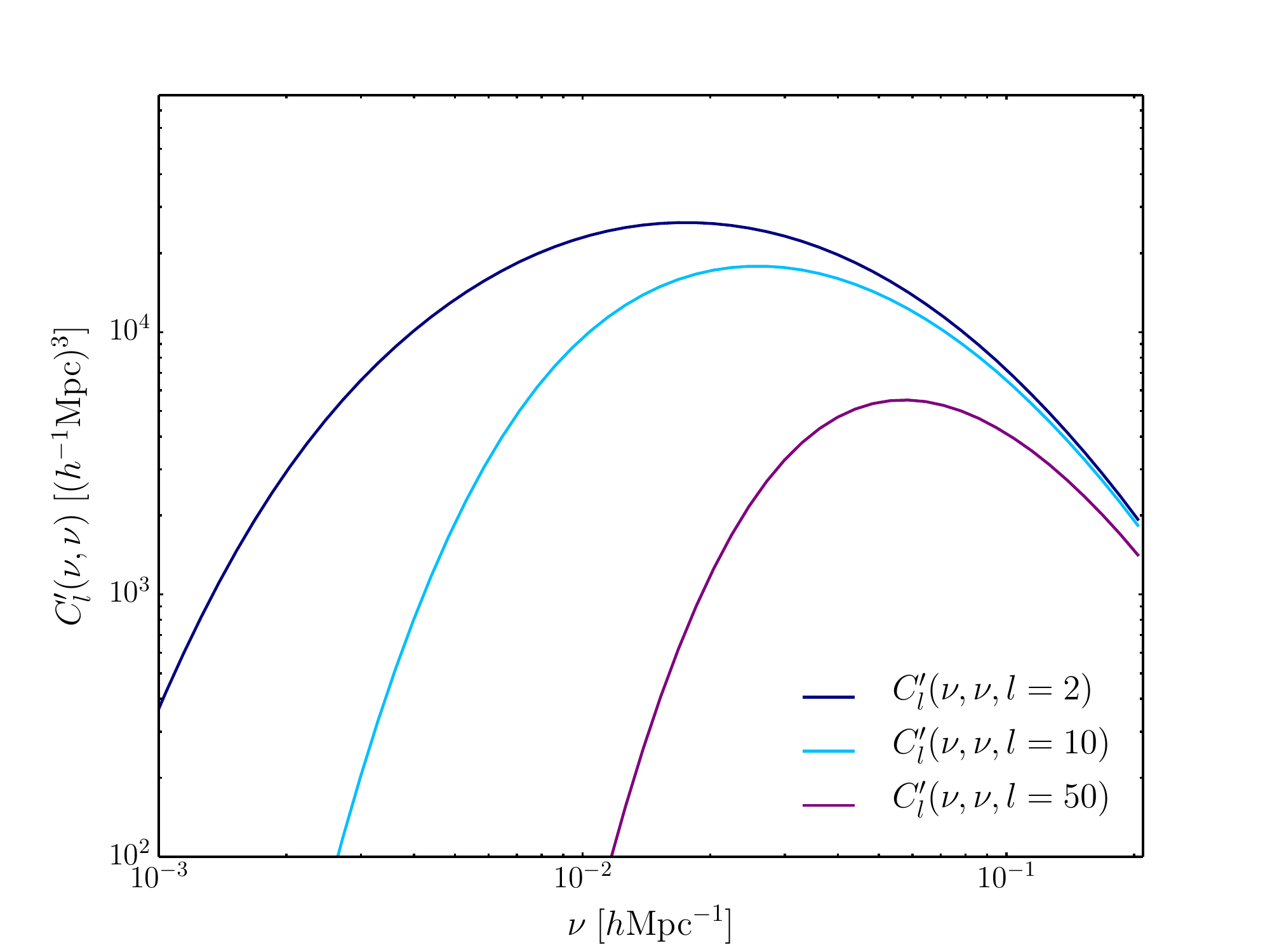}} 
\caption{The SFB auto power spectrum $C'_{l} (\nu, \nu)$ for two different distance redshift relations as a function of angular scale $l$.} 
\label{fig:Cls_SFB_dist}
\end{center}
\end{figure}

\section{\label{ap:FisherMatrices} Derivation of the Fisher matrix for the SFB power spectrum}

\subsection{\label{subap:Discrete_Fisher} Full Fisher matrix}

The FM obtained from a data likelihood with covariance matrix $\textbf{C}$ and mean $\boldsymbol \mu$ is given by \cite{1997ApJ...480...22T}
\begin{equation}
F_{\alpha \beta} =  \frac{1}{2} \Tr \left [ \textbf{C}^{-1} \frac{\partial \textbf{C}}{\partial \theta_{\alpha}} \textbf{C}^{-1} \frac{\partial \textbf{C}}{\partial \theta_{\beta}} \right ] + \frac{\partial \boldsymbol \mu^{T}}{\partial \theta_{\alpha}} \textbf{C}^{-1} \frac{\partial \boldsymbol \mu}{\partial \theta_{\beta}} 
\label{eq:FisherMat_exp}
\end{equation}
Assuming a measurement of the SFB coefficients of the matter overdensity field $\delta_{lm}(k)$ for a discrete set of radial wave vectors $k$ denoted $(k_{1}, k_{2}, \cdots, k_{n})$, the data covariance matrix is given by 
\begin{equation}
\hat{\textbf{C}} =
 \begin{pmatrix}
  \hat{\textbf{C}}_{l_{1}} & 0 & 0 & 0 & \cdots & 0 \\
  \vdots  & \ddots  & \vdots & \vdots & \ddots & \vdots  \\
    0 & \cdots & \hat{\textbf{C}}_{l_{1}} & 0 & \cdots & 0 \\
    0 & 0 & 0 & \hat{\textbf{C}}_{l_{2}} & \cdots & 0 \\
    \vdots  & \vdots  & \vdots & \vdots & \ddots & \vdots  \\
     0 & 0 & 0 & 0 & \cdots & \hat{\textbf{C}}_{l_{max}} & 
 \end{pmatrix}
\end{equation}
where the sub-covariance matrices are defined as
\begin{equation}
\hat{\textbf{C}}_{l_{i}} =
 \begin{pmatrix}
  \tilde{C}_{l_{i}} (k_{1}, k_{1}) & \tilde{C}_{l_{i}} (k_{1}, k_{2}) & \cdots & \tilde{C}_{l_{i}} (k_{1}, k_{n}) \\
  \tilde{C}_{l_{i}} (k_{2}, k_{1}) & \tilde{C}_{l_{i}} (k_{2}, k_{2}) & \cdots & \tilde{C}_{l_{i}} (k_{2}, k_{n}) \\
  \vdots  & \vdots  & \ddots & \vdots  \\
  \tilde{C}_{l_{i}} (k_{n}, k_{1}) & \tilde{C}_{l_{i}} (k_{n}, k_{2}) & \cdots & \tilde{C}_{l_{i}} (k_{n}, k_{n})
 \end{pmatrix}
\end{equation}
and $\tilde{C}_{l}(k_{i}, k_{j}) = C_{l}(k_{i}, k_{j}) + N_{l}(k_{i}, k_{j})$. Since $\hat{\textbf{C}}$ is blockdiagonal and $\mu = \langle \delta_{lm}(k_{i}) \rangle = 0 $, Eq.~\ref{eq:FisherMat_exp} yields 
\begin{equation}
F_{\alpha \beta} = f_{\text{sky}} \sum_{l} \frac{\left (2 l +1 \right ) \Delta l}{2} \Tr \left [ \hat{\textbf{C}}_{l}^{-1} \frac{\partial \hat{\textbf{C}}_{l}}{\partial \theta_{\alpha}} \hat{\textbf{C}}_{l}^{-1} \frac{\partial \hat{\textbf{C}}_{l}}{\partial \theta_{\beta}} \right ]
\end{equation}
where the sum is over bands of width $\Delta l$ in the power spectrum.

\subsection{\label{subap:Continuous_Fisher} Diagonal Fisher matrix}

The computation of the SFB FM through Eq.~\ref{eq:Fisher_disc} is numerically challenging and it is therefore desirable to investigate possible approximations. Although the amount of cross-correlation $C_{l}(k, k')$ between neighbouring $k$ vectors can be considerable, it tends rapidly to zero for separated wave vectors. If we assume a broad window function, such that mode coupling can be neglected \cite{1994ApJ...426...23F}, we can approximate $C_{l}(k, k') = 0 $  for $k \neq k'$ \footnote{This approximation is equally justified when we assume that the SFB power spectrum is computed for radial wave vector bins which are broader than the correlation scale due to finite survey effects.}. Assuming a measurement of a set of discrete SFB modes $\delta_{l m}(k_{i})$ up to $l \leq l_{\text{max}}$which satisfy $\langle \delta_{l m}(k_{i}) \rangle = 0$ and defining
\begin{eqnarray}
&& \delta_{l m}(k_{i}) = \delta_{l m, i} \\
&& \langle \delta_{l m}(k_{i})  \delta_{l m}(k_{i}) \rangle = C_{l} (k_{i}, k_{i}) + N_{l}(k_{i}, k_{i}) = \Delta^{2}_{l, i} \nonumber 
\end{eqnarray}
the data likelihood can be written as
 \begin{equation}
L \left (\textbf{x}; \boldsymbol \theta \right ) = \frac{1}{(2 \pi)^{l_{\text{max}} (l_{\text{max}} + 2) \frac{n}{2}} \prod_{l, i} \Delta_{l, i}^{2l +1}} e^{-\frac{1}{2} \sum_{l, m, i} \frac{\delta^{2}_{l m, i}}{\Delta^{2}_{l, i}}}
\end{equation} 
Applying Eq.~\ref{eq:FisherMat}, the FM becomes
\begin{equation}
F_{\alpha \beta} = f_{\text{sky}} \sum_{l, i} \frac{\left (2 l +1 \right ) \Delta l}{2} \frac{1}{(C_{l} (k_{i}, k_{i}) + N_{l}(k_{i}, k_{i}))^{2}} \frac{\partial C_{l} (k_{i}, k_{i})}{\partial \theta_{\alpha}}\frac{\partial C_{l} (k_{i}, k_{i})}{\partial \theta_{\beta}}
\label{eq:FisherMat_RS}
\end{equation}
where the sum is over bands of width $\Delta l$ in the power spectrum and wave vectors $k_{i}$. To proceed, we assume that the maximal length scale probed by the survey is given by $L$. Therefore the minimal measurable mode is $k_{\text{min}} = \sfrac{2 \pi}{L}$ which also defines the k-space resolution. The maximal measurable mode is determined by the smallest distance $\Delta L$ and given by $k_{\text{max}} = \sfrac{2 \pi}{\Delta L}$. Turning the Riemann sum in Eq~\ref{eq:FisherMat_RS} into a continuous integral yields
\begin{equation}
F_{\alpha \beta} = f_{\text{sky}} \sum_{l} \frac{\left (2 l +1 \right ) \Delta l}{2} \int \limits_{k_{\text{min}}}^{k_{\text{max}}} \frac{L \text{d} k}{2 \pi} \frac{1}{\left ( C_{l}(k,k) + N_{l}(k,k) \right )^{2}} \frac{\partial C_{l}(k,k)}{\partial \theta_{\alpha}} \frac{\partial C_{l}(k,k)}{\partial \theta_{\beta}}
\end{equation}

\section{\label{ap:TM} Toy models illustrating effects of redshift binning}

\subsection{\label{ap:TMI} Toy model I}

Investigating the exact cause for changes in cosmological Fisher matrix calculations due to different redshift binning schemes in tomographic analyses is complicated by the large number of cosmological parameters to constrain and the degeneracies between those. To study these we therefore resort to highly simplified toy models, which are designed to be mostly analytically solvable. We believe that such a simplified treatment allows us to interpret results more easily.

We assume a toy model matter power spectrum, defined as
\begin{equation}
P(k, r) = A_{0} \left (\frac{r}{r_{0}} \right )^{-\beta} \left (\frac{k}{k_{0}} \right )^{-\alpha}
\label{eq:MockPowerSpec}
\end{equation}
where $k$ is the wave vector, $r$ the comoving distance and $k_{0}$, $r_{0}$ are normalisation constants, assumed to be precisely known. The power spectrum is specified by the parameters $\alpha$, $\beta$ and $A_{0}$, which mimic the three generic features of the linear $\Lambda \text{CDM}$ matter power spectrum: $A_{0}$ is a multiplicative amplitude, $\alpha$ imitates the spectral index and $\beta$ determines the growth of structure. In accordance with the cosmological power spectrum we set their fiducial values to $A_{0} = 1$, $\alpha = 3$ and $\beta = 1$, where the last two equalities mimic the slope of the high k matter power spectrum and the growth factor in a matter dominated universe. 

\begin{figure}
\begin{center}
\subfigure[Two bins]{\includegraphics[width=0.49\textwidth]{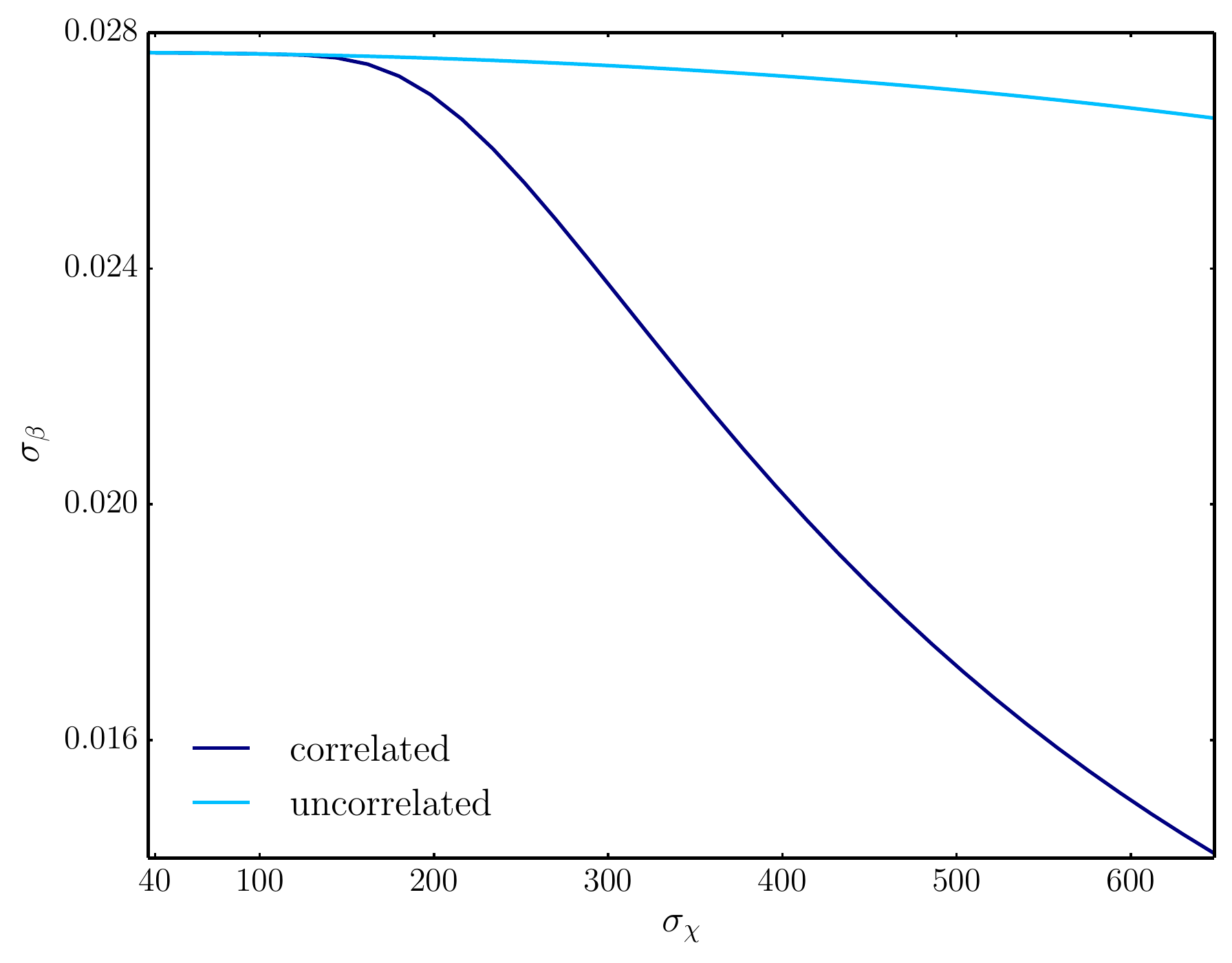}}
\subfigure[Several bins]{\includegraphics[width=0.49\textwidth]{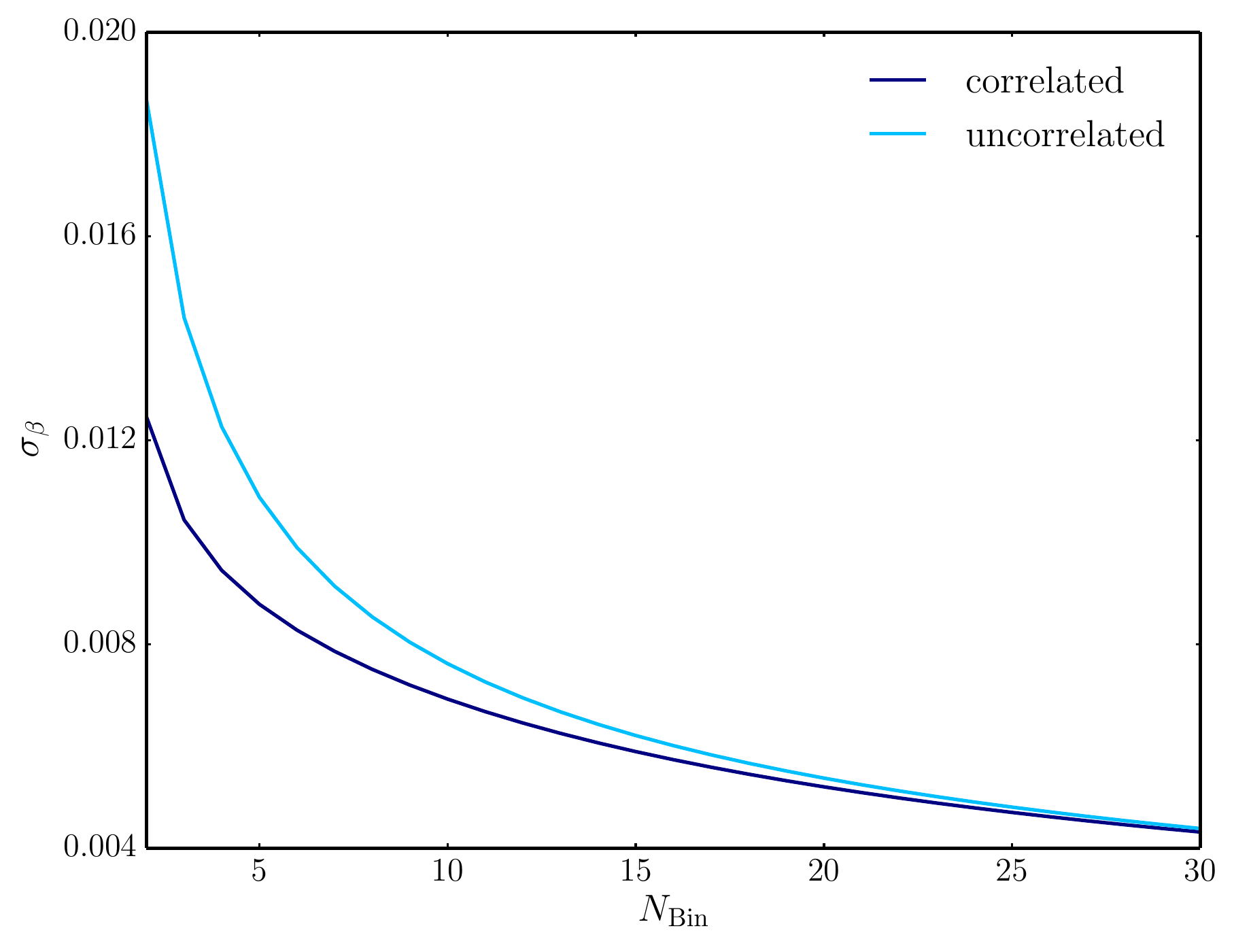}} 
\caption{The marginalised uncertainty on $\beta$ as a function of bin variance i.e. overlap on the left as well as bin number on the right.} 
\label{fig:TM1}
\end{center}
\end{figure}

We consider tomographic analyses of a galaxy redshift survey with varying bin geometries. Because the qualitative results of the toy model do not depend on the order of accuracy of the Limber approximation, we compute the spherical harmonic tomography power spectrum employing a simplified version. This approximation yields results, which are easier to interpret, and is given by
\begin{equation} 
C_{l, \text{Limber}} = \int \text{d} r \frac{\phi_{i}(r) \phi_{j}(r)}{r^{2}} P \left (k = \frac{l}{r}, r \right ) 
\label{eq:ClLimber_chi}
\end{equation} 
where $i$ and $j$ label the selection function of a particular redshift bin. We consider two Gaussian bins, each characterised by its mean $\bar{r}_{i}$ and variance $\sigma_{i}$, which are both accurately known. In this minimal tomographic analysis, we vary the amount of overlap between the two bins by increasing their variances while keeping their means fixed. For each bin configuration we compute the parameter constraints using Eq.~\ref{eq:FisherClz} once assuming physically distinct, i.e. uncorrelated redshift bins and once assuming them to be correlated (implementation details are described in Table \ref{tab:toymodels}) \footnote{In practice we set the cross-correlation between the bins to zero in the first case, while in the second case we take it into account to determine parameter constraints.}. These two settings correspond to conducting a survey in two different parts of the sky or on the same sky patch respectively. To further simplify calculations, we assume the total number of surveyed galaxies to be large so that measurement uncertainties due to shot noise can be neglected, meaning that our conclusions are only applicable to non-shot noise dominated surveys. Fig.~\ref{fig:TM1} shows the constraints on $\beta$ as a function of bin variance for constant bin mean separation $\Delta r \simeq 500 \: \text{Mpc}$. The behaviour suggests that configurations with more correlation give tighter constraints on $\beta$, because the constraints for correlated bins decrease with increasing bin overlap while those for uncorrelated bins stay approximately constant. The constraints on $A_{0}$ exhibit an analogous behaviour, while those on $\alpha$, as well as the fixed parameter constraints, depend only weakly on overlap and thus correlation. From Eq.~\ref{eq:MockPowerSpec} we see that while $A_{0}$ and $\beta$ exhibit considerable redshift degeneracies, $\alpha$ is the only parameter which affects the wave vector dependence of the power spectrum and it is not degenerate with the others. This suggests that increased bin overlap mainly introduces correlations between bins which help breaking redshift degeneracies between parameters.

As the number of redshift bins used in the tomographic analysis is gradually increased from $N_{\text{Bin}} = 2$ to $N_{\text{Bin}} = 30$, bin overlap becomes increasingly less important as shown in Fig.~\ref{fig:TM1}. This suggests that the information gained from the cross-correlation between overlapping bins becomes negligible as we recover an increasing amount of information from the survey.

\begin{table}
\caption{Specification of Toy models} \label{tab:toymodels}
\begin{center}
\begin{tabular}{>{\centering}m{2.5cm}|>{\centering}m{10cm}@{}m{0pt}@{}} \hline \hline
                                 
Toy model I & 
\multicolumn{2}{>{\centering}m{10cm}}{Sky coverage: $f_{\text{sky}} = 0.125$ \\
Angular scales covered: $l \in [1, 1000]$ \\
Parameters: $\boldsymbol{\theta} = (\alpha, \beta, A_{0})$ \\
Fiducial values: $\boldsymbol{\theta}_{\text{fid}} = (3.0, 1.0, 1.0)$  \\
Two bins: $\bar{r}_{1} = 2116. \: h^{-1} \text{Mpc}$, $\bar{r}_{2} = 2607. \: h^{-1} \text{Mpc}$ \\
Several bins: means equally spaced in $[1656., 3068.] \: h^{-1} \text{Mpc}$} 

\\ \hline 
Toy model II & 
\multicolumn{2}{>{\centering}m{10cm}}{Parameters: $\boldsymbol{\theta} = (\theta_{1}, \theta_{2})$ \\
Fiducial values: none \\
Measurement points (arbitrary units): $x_{1} = 1$, $x_{2} = 2$ \\
Measurement uncertainties (arbitrary units): $\sigma_{1} = 0.1$, $\sigma_{2} = 0.1$} 

\\ \hline \hline

\end{tabular}
\end{center}
\end{table} 

\subsection{\label{ap:TMII} Toy model II}

The results from toy model I suggest that parameter constraints can depend on the amount of correlation between redshift bins, which results in a dependence of Fisher matrix constraints on binning scheme for small bin numbers. In order to understand the reasons for these results we resort to an even simpler model.

Toy model I is in its essence identical to the problem of fitting a straight line through two data points with correlated errors. We can thus try to gain intuition about the former by considering this trivial problem.

We assume conducting two measurements of a given physical quantity $y$ at the points $x_{1}$ and $x_{2}$. The measured values are denoted $y_{1}$, $y_{2}$ and the correlation between these data points is allowed to vary from no correlation to full positive or negative correlation. Mathematically we can describe this situation by assuming that the two data points $y_{1}$ and $y_{2}$ follow a bivariate Gaussian probability distribution given by 
\begin{equation}
L \left ( \textbf{y} | \boldsymbol{\theta} \right ) = \frac{1}{2 \pi \sqrt{\det \textbf{C}}} e^{- \frac{1}{2} \left ( \textbf{y} - \bar{\textbf{y}} \right ) \textbf{C}^{-1} \left ( \textbf{y} - \bar{\textbf{y}} \right )}
\label{eq:datalikelihood}
\end{equation} 
where the means of the distribution are assumed to linearly depend on the measurement points through the parameters $\theta_{1}$, $\theta_{2}$ i.e.
\begin{equation}
\bar{y}_{i} = \theta_{1} x_{i} + \theta_{2}
\end{equation} 
The data covariance matrix can be written as 
\begin{equation}
\textbf{C} = \begin{pmatrix}
 \sigma_{1}^{2} & r \sigma_{1} \sigma_{2} \\
 r \sigma_{1} \sigma_{2} & \sigma_{2}^{2} \\
\end{pmatrix}
\label{eq:data_covmatrix}
\end{equation}
where $\sigma_{i}$ denotes the respective variance in the measurement and the correlation between the two data points is quantified by the correlation coefficient $r$ with $|r| \le 1$, which is defined by \cite{Press2007}
\begin{equation}
r = \frac{\left \langle \left ( y_{1} - \bar{y}_{1} \right ) \left ( y_{2} - \bar{y}_{2} \right ) \right \rangle}{\sigma_{1} \sigma_{2}}
\label{eq:corrcoeff}
\end{equation} 
We can compute the conditional probabilities for both variables i.e. the probability distribution of the second measurement after having conducted the first. As an example, the conditional probability for measurement $y_{2}$ given measurement $y_{1}$ is
\begin{align} \nonumber
L \left ( y_{2} | \boldsymbol{\theta}, y_{1} \right ) &= \frac{L \left ( \textbf{y} | \boldsymbol{\theta} \right )}{L \left ( y_{1} | \boldsymbol{\theta} \right )} \\
&= \frac{1}{\sqrt{2 \pi} \sigma_{2} \sqrt{1-r^{2}}} e^{- \frac{1}{2 \sigma^{2}_{2} ( 1-r^{2} )} \left [ y_{2} - \bar{y}_{2} -r \frac{\sigma_{2}}{\sigma_{1}} \left ( y_{1} - \bar{y}_{1} \right ) \right ]^{2}}
\end{align}  
This is also a Gaussian with mean and variance
\begin{align}
\text{E} \left ( y_{2} | y_{1} \right ) &=  \bar{y}_{2} + r \frac{\sigma_{2}}{\sigma_{1}} \left ( y_{1} - \bar{y}_{1} \right ) \label{eq:meany2} \\ 
\text{Var} \left ( y_{2} | y_{1} \right ) &=  \sigma^{2}_{2} ( 1-r^{2} ) \label{eq:vary2}
\end{align}
When the two measurements are correlated, the separation of the first measurement from its mean determines that of the second one from its respective mean. In the case of positive nonzero correlation the measurements will therefore both lie either above or below their respective means. For negative nonzero correlation on the other hand, one measurement will tend to overestimate while the other will tend to underestimate its respective mean. A correlation between measurements thus provides information on the relative location of the data points. Constraints on one particular data point do not benefit from this kind of information, while constraints on any combination of data points on the other hand will be sensitive to it.

To investigate how correlations between data points affect constraints on the straight line parameters $\theta_{1}$ and $\theta_{2}$, we compute their forecasted FM uncertainties assuming a flat prior from Eq.~\ref{eq:FisherMat} (implementation details are described in Table \ref{tab:toymodels}). 

\begin{figure}
\begin{center}
\includegraphics[scale=0.5]{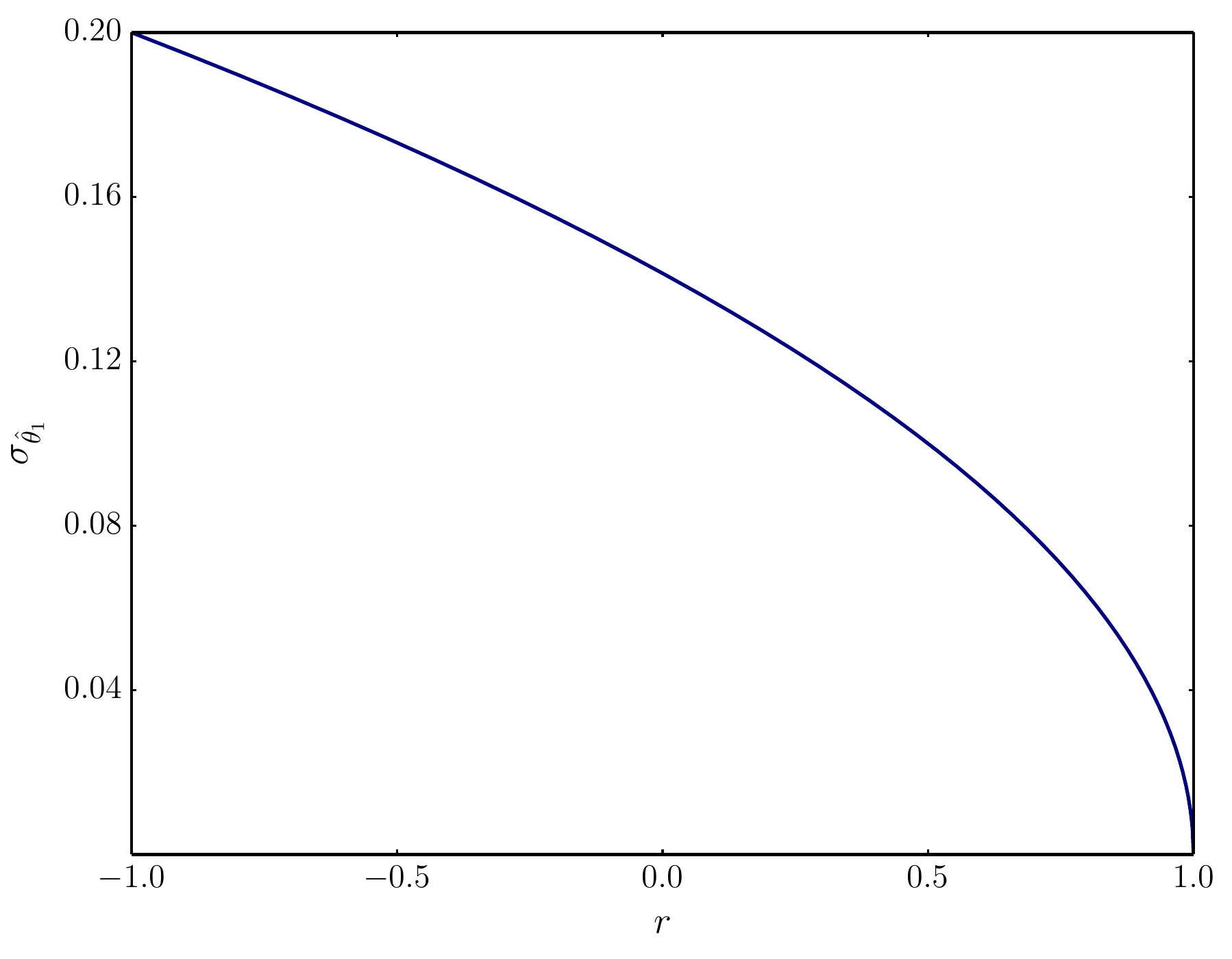}
\caption{The marginalised uncertainty on the slope $\hat{\theta}_{1}$ as a function of correlation coefficient $r$.} 
\label{fig:TM2}
\end{center}
\end{figure}

Fig.~\ref{fig:TM2} shows the constraints on the slope $\theta_{1}$ as a function of correlation coefficient $r$. As the correlation between the data points is increased, the constraints improve, a behaviour analogous to that of TM I. Being fully analytic, straight line fitting allows us to examine the expression for the maximum likelihood estimator (MLE) for the slope, which is given by
\begin{equation}
\hat{\theta}_{1} = \frac{y_{2} - y_{1}}{x_{2} - x_{1}} 
\label{eq:slope_comb} 
\end{equation}
This quantity directly depends on the difference between $y_{1}$ and $y_{2}$. When these two data points are positively correlated they both lie either above or below their respective means, implying that the errors will tend to cancel when computing the uncertainty on $\hat{\theta}_{1}$. The opposite applies to the sum of the two data points, since in this case the errors for positive correlation will tend to add. As $r$ is increased, Equations \ref{eq:meany2} and \ref{eq:vary2} show that the second measurement increasingly depends on the first i.e. its independent variance decreases \cite{Fisher1970}, an effect which further reduces uncertainties on the difference of data points.

These trivial considerations therefore suggest that constraints on parameters, which depend on combinations of the data, are sensitive to correlations, because those provide information about the relative location of data points. Applying this to tomographic analyses of galaxy redshift surveys suggests that cosmological parameters can be divided into two classes, depending on how constraints depend on correlations: (i) Redshift degenerate parameters can only be simultaneously constrained with redshift leverage. The estimators for these parameters are likely to depend on a combination of the data used to constrain them, making their uncertainties sensitive to correlations. Furthermore if the parameters happen to depend on the difference between the various measurements, the uncertainties will tend to decrease as we increase the amount of correlation, as is found in the calculations of \ref{ap:TMI}. (ii) Parameters that can be distinguished from all others on the other hand can already be constrained with only one data point. Their maximum likelihood estimator will likely only depend on one data point and therefore the constraints on such non-redshift degenerate parameters are expected to show a weak dependence (if any) on the amount of correlation between the data.

Since the amount of overlap between redshift bins used in a tomographic analysis of galaxy redshift surveys affects the level of correlation between the data, the observed dependence of parameter constraints on binning scheme is probably due to the ``directional information'' contained in the correlations. The counter-intuitive overlap dependence of parameter constraints is therefore probably a manifestation of a generic feature of correlated data sets.

\section{\label{ap:results_FMs} Comparison of FM computation techniques for the SFB power spectrum: Results}

The parameter constraints obtained for the SFB power spectrum using both Fisher matrix (FM) computation methods and applying the same wave vector cuts, both neglecting and including shot noise contributions, are shown in Table \ref{tab:full_vs_approx}. Apart from the constraints on $w_{0}$ the results agree reasonably well, as mentioned in \ref{subsubsec:FMComparisons}.

\begin{table}[h]
\caption{Comparison of parameter constraints for the SFB power spectrum obtained using discrete (Eq.~\ref{eq:Fisher_disc}) and continuous (Eq.~\ref{eq:Fisher_cont}) Fisher matrix} \label{tab:full_vs_approx}
\begin{center}
\begin{ruledtabular}
\begin{tabular}{cccccccccc}
Statistic & \multicolumn{2}{c}{Implementation} &Shot noise & $\sigma_{h}$ & $\sigma_{\Omega_{\text{m}}}$ & $\sigma_{\Omega_{\Lambda}}$ &  $\sigma_{w_{0}}$ & $\sigma_{n_{\text{s}}}$ & $\sigma_{\sigma_{8}}$ \\ \hline
                                
\multirow{4}{*}{SFB} & \multirow{2}{*}{full cov.} & Comoving & no & 0.38 & 0.043 & 0.55 & 1.3 & 0.38 & 0.86 \\ 
& & Comoving & yes & 0.48 & 0.056 & 0.68 & 1.5 & 0.46 & 1.1 \\ \cline{2-10}
& \multirow{2}{*}{diag cov.} & Comoving & no & 0.30 & 0.071 & 0.57 & 2.2 & 0.27 & 0.66  \\ 
& & Comoving & yes & 0.39 & 0.10 & 0.77 & 3.0 & 0.33 & 0.85  \\ 
\end{tabular}
\end{ruledtabular}

\end{center}
\end{table} 

\bibliography{Bib3D_analyses}

\end{document}